\documentclass[twocolumn,tighten]{aastex631}
\usepackage{xspace}
\usepackage{eucal}
\usepackage{float}
\usepackage{graphicx} 
\usepackage{amsmath, esint}
\usepackage{amssymb}

\newcommand{\rhigh}{$R_{\text{high}}$ }

\def\lsim{\mathrel{\raise.3ex\hbox{$<$\kern-.75em\lower1ex\hbox{$\sim$}}}}
\def\gsim{\mathrel{\raise.3ex\hbox{$>$\kern-.75em\lower1ex\hbox{$\sim$}}}}

\defcitealias{EHTC_M87_I}{\m87~I}
\defcitealias{EHTC_M87_II}{\m87~II}
\defcitealias{EHTC_M87_III}{\m87~III}
\defcitealias{EHTC_M87_IV}{\m87~IV}
\defcitealias{EHTC_M87_V}{\m87~V}
\defcitealias{EHTC_M87_VI}{\m87~VI}
\defcitealias{EHTC_M87_VII}{\m87~VII}
\defcitealias{EHTC_M87_VIII}{\m87~VIII}
\defcitealias{EHTC_M87_IX}{\m87~IX}
\defcitealias{EHTC_M87_2018}{\m87-2018-I}
\defcitealias{EHTC_SgrA_I}{\sgra~I}
\defcitealias{EHTC_SgrA_II}{\sgra~II}
\defcitealias{EHTC_SgrA_III}{\sgra~III}
\defcitealias{EHTC_SgrA_IV}{\sgra~IV}
\defcitealias{EHTC_SgrA_V}{\sgra~V}
\defcitealias{EHTC_SgrA_VI}{\sgra~VI}
\defcitealias{EHTC_SgrA_VII}{\sgra~VII}
\defcitealias{EHTC_SgrA_VIII}{\sgra~VIII}

\def\m87{M87$^*$\xspace}
\def\sgra{Sgr~A$^*$\xspace}

\begin{document}
\title{Bayesian Black Hole Photogrammetry}
\correspondingauthor{Dominic O. Chang}
\email{dochang@g.harvard.edu}

\author[0000-0001-9939-5257]{Dominic O. Chang}
\affiliation{Department of Physics, Harvard University, Cambridge, Massachusetts 02138, USA}
\affiliation{Black Hole Initiative at Harvard University, 20 Garden Street, Cambridge, MA 02138, USA}

\author[0000-0002-4120-3029]{Michael D. Johnson}
\affiliation{Black Hole Initiative at Harvard University, 20 Garden Street, Cambridge, MA 02138, USA}
\affiliation{Center for Astrophysics $|$ Harvard \& Smithsonian, 60 Garden Street, Cambridge, MA 02138, USA}

\author[0000-0003-3826-5648]{Paul Tiede}
\affiliation{Black Hole Initiative at Harvard University, 20 Garden Street, Cambridge, MA 02138, USA}
\affiliation{Center for Astrophysics $|$ Harvard \& Smithsonian, 60 Garden Street, Cambridge, MA 02138, USA}

\author[0000-0002-7179-3816]{Daniel~C.~M.~Palumbo}
\affiliation{Black Hole Initiative at Harvard University, 20 Garden Street, Cambridge, MA 02138, USA}
\affiliation{Center for Astrophysics $|$ Harvard \& Smithsonian, 60 Garden Street, Cambridge, MA 02138, USA}


\begin{abstract}
We propose a simple, analytic dual-cone accretion model for horizon scale images of the cores of Low-Luminosity Active Galactic Nuclei (LLAGN), including those observed by the Event Horizon Telescope (EHT). 
Our underlying model is of synchrotron emission from an axisymmetric, magnetized plasma, which is constrained to flow within two oppositely oriented cones that are aligned with the black hole's spin axis. 
We show that this model can accurately reproduce images for a variety of time-averaged general relativistic magnetohydrodynamic (GRMHD) simulations, that it accurately recovers both the black hole and emission parameters from these simulations, and that it is sufficiently efficient to be used to measure these parameters in a Bayesian inference framework with radio interferometric data. 
We show that non-trivial topologies in the source image can result in non-trivial multi-modal solutions when applied to observations from a sparse array, such as the EHT 2017 observations of \m87.  
The presence of these degeneracies underscores the importance of employing Bayesian techniques that adequately sample the posterior space for the interpretation of EHT measurements. 
We fit our model to the EHT observations of \m87 and find a 95\% Highest Posterior Density Interval (HPDI) for the mass-to-distance ratio of $\theta_g\in(2.84,3.75)\,\mu{\rm as}$, and give an inclination of $\theta_{\rm o}\in(11^\circ,24^\circ)$. These new measurements are consistent with mass measurements from the EHT and stellar dynamical estimates \citep[e.g.,][]{Gebhardt,EHTC_M87_IV,EHTC_M87_VI,Liepold_2023}, and with the spin axis inclination inferred from properties of the \m87 jet \citep[e.g.,][]{Walker}. 
\end{abstract}

\section{Introduction}
\label{sec:introduction}
In 2017, the Event Horizon Telescope Collaboration (EHTC) observed the supermassive black hole, \m87, at the center of the elliptical galaxy, Messier 87 using a 7-element Very Long Baseline Interferometry (VLBI) array operating at $\lambda=1.3\,{\rm mm}$. 
The resulting EHT images of \m87 revealed a dark central brightness depression, consistent with expectations for accretion flows within a Kerr spacetime \citep[][hereafter \m87~I-IX]{EHTC_M87_I,EHTC_M87_II,EHTC_M87_III,EHTC_M87_IV,EHTC_M87_V,EHTC_M87_VI,EHTC_M87_VII,EHTC_M87_VIII,EHTC_M87_IX}. More recently, the EHTC has also published images of the supermassive black hole \sgra at the center of the Milky Way \citep[][hereafter \sgra~I-VIII]{EHTC_SgrA_I,EHTC_SgrA_II,EHTC_SgrA_III,EHTC_SgrA_IV,EHTC_SgrA_V,EHTC_SgrA_VI,EHTC_SgrA_VII,EHTC_SgrA_VIII}.
These images now provide the most direct evidence for the existence of supermassive black holes. 

A crucial question is what insights into the properties of both supermassive black holes and their immediate environments are possible through the EHT measurements, a question that falls under the broader science of \emph{photogrammetry}. 
For instance, the 2017 campaign allowed the EHTC to constrain the diameter of the bright emission ring in \m87 to be $d = 42\pm 3\, \mu\text{as}$.
General Relativity predicts a direct relationship between the mass-to-distance ratio and spin of the black hole and its shadow diameter and shape \citep{Bardeen_1973}. 
However, mass measurements from the diameter of the emission ring on an image (often called the diameter of the ``apparent shadow'') are dependent on assumptions about the emission geometry (see, for example, \citetalias{EHTC_M87_V}, \citetalias{EHTC_M87_VI}, and \citealt{Gralla_2019}). 
To relate the measured diameter of the emission ring to the black hole mass-to-distance ratio (or angular gravitational radius), $\theta_g \equiv GM/(c^2 D)$, the EHTC introduced an unknown scaling factor, $\alpha$ \citepalias{EHTC_M87_VI}: 
\begin{align}
    d = \alpha \theta_g.
\end{align}
The EHTC estimated $\alpha$ and its associated uncertainty using a suite of general relativistic magnetohydrodynamic (GRMHD) simulations \citepalias{EHTC_M87_V}. 
These ab initio simulations self-consistently evolve a magnetized fluid in the Kerr spacetime and have been successful in producing a broad range of phenomena seen in LLAGN, from horizon-scale image structures and variability seen with the EHT to powerful, large-scale jets with multi-wavelength emission. 
This approach resulted in a reported mass measurement for \m87 of $M=6.5\pm0.2\rvert_{\text{stat}}\pm0.7\rvert_{\text{sys}}\times 10^9 M\textsubscript{\(\odot\)}$, \citepalias{EHTC_M87_VI}. In this estimate, the dominant uncertainty corresponds to that of $\alpha$ (i.e., systematic uncertainty associated with the unknown emission properties and spin).

Apart from the EHTC's mass measurement, there is a rich literature of other methods to infer the mass of \m87 using observations on much larger scales. 
These methods model the resolved emission spectra of \m87 originating from different components, usually either from the dynamics of gas in the circumnuclear disk around \m87 \citep[see, for example,][]{Macchetto_1997,Walsh} or from the stellar velocity dispersion in the bulge \citep[$\sigma_*$; see][]{Gebhardt, Liepold_2023,Simon}. 
There has historically been a discrepancy in the measured mass, both between and within these two methodologies. 
The gas dynamical mass measurements are typically only ${\sim}$half the value of the smallest stellar dynamics measurements. Moreover, different modelling choices may result in large systematic uncertainties \citep[see, e.g.,][]{Jeter_2021,Simon}. 

The EHTC results provide a crucial input to assess these alternative mass measurements, but the $\alpha$-calibration approach may also have unknown systematic errors and is dependent on the selection of GRMHD models \citepalias{EHTC_M87_V}. More recent analyses using semi-analytic models have found that this type of calibration can provide estimates for the mass-to-distance ratio with ${\sim}10\%$ systematic uncertainty \citep{Ozel_2022,Younsi_2023,EHTC_SgrA_VI}. Nevertheless, even for GRMHD simulations, some models in the EHTC library would give significantly different mass inferences but were excluded based on additional constraints on the observed jet power and x-ray luminosity of \m87 \citepalias{EHTC_M87_V}. Hence, other approaches to black hole parameter inference from EHT data are of significant value in substantiating these conclusions and could provide sharper estimates of the black hole properties.



\citet{Palumbo_2022} provided one such alternative: a mass inference scheme that involves fitting EHT data using a model whose parameters are described by both the black hole spacetime and a flexible emission geometry. 
The emission model was taken to be purely equatorial and optically thin. 
Using time-averaged GRMHD simulations, the authors show that this approach gives accurate mass measurements for magnetically arrested disk \citep[MAD;][]{Narayan_2003} simulations, albeit with a modest (${\sim}10\%$) underestimate in some simulations. 
However, the approach gives poorer mass estimates for standard and normal evolution \citep[SANE;][]{Narayan_2012} simulations. 
Although the authors did not infer a mass from the true EHT data, which favor MAD models, their work shows the potential for using simple emission models to derive information about a black hole mass and emission geometry from EHT data.

In this paper, we develop a simple, flexible model for images of accretion flows around supermassive black holes.
We show that the model is capable of not only inferring spacetime parameters, but also measuring the 3D structure of the inner emission geometry---effectively enabling photogrammetry within a curved spacetime.
Our model extends the equatorial version defined in \citet{Palumbo_2022} to a bi-conical region aligned with the spin axis, motivated by optically thin emitting material concentrated within a jet sheath or accretion disk near a Kerr black hole as seen in a variety of numerical simulations \citep[e.g.,][]{Dexter,Moscibrodzka,EHTC_M87_V,Wong_2021}. 
The primary advantage of our model relative to more sophisticated physical approaches is that it is analytic and differentiable.
These features allow the model to be suitable for performing Bayesian inference of black hole spacetime and emission parameters directly from interferometric measurements, such as those of the EHT. 
In addition, the model captures the salient features in GRMHD images for both MAD and SANE simulations and allows a crisp exploration of degeneracies in images that can arise between the parameters of the black hole and those of the synchrotron-emitting plasma.

The paper is outlined as follows.
In \autoref{sec:theory}, we introduce our emission model and describe how it is used to produce images. 
In \autoref{sec:image_domain}, we assess the ability of this model to reproduce the images seen in a variety of GRMHD simulations. 
In \autoref{sec:visibility_domain}, we use this model to fit EHT observations of \m87. 
In \autoref{sec:discussion}, we summarize our results.\footnote{A repository of scripts used in this analysis are open source and are available on  \href{https://github.com/dominic-chang/BayesianBlackHolePhotogrammetry}{\textcolor{blue}{\underline{github}}}.}

\section{Description of the model}
\label{sec:theory}
\subsection{Motivation \& Assumptions}
\label{sec:existing_models}

Many authors have previously used simple geometric models of the inner accretion flow of supermassive black holes to extract space time parameters from very long baseline (VLBI) data.
These efforts include a series of works that developed semi-analytic models of a radiatively inefficient accretion flow \citep[RIAF;][]{Broderick_2009,Broderick_2011, Broderick_2014,Broderick_2016,Pu_2016,Pu_2018,Broderick_2020} and of emission from jets \citep[e.g.,][]{Broderick_2009b}. Those works report constraints on the spin, inclination and position angle of \m87 and \sgra.

These simple models complement the more complex GRMHD simulations that are the basis for most theoretical interpretations of the EHTC. Both types of models have been used to show that some image features, such as the black hole's apparent shadow, are largely independent of the underlying accretion physics when observed at $mm$ wavelengths (for example, the conservative fluid models of \citet{Ozel, Ziri, Papoutsis} and general relativistic particle in cell (GRPIC) models of \citet{Galishnikova_2023} yield similar results). GRMHD simulations have the additional benefit of self-consistently modeling source variability. The 2017 observations of \sgra by the EHT, however imply that the observed variability is in tension with the predictions of GRMHD \citepalias[see, e.g.,][]{EHTC_SgrA_IV}.


Our ignorance of the precise geometry and distribution function of the emitting plasma motivates us to develop a model that is flexible enough to account for variations seen in a variety of accretion scenarios, while still being constraining enough to give meaningful information about the black hole and its accretion system even with sparse interferometric data. In creating this model, we are guided by the time-averaged emission structure from GRMHD simulations. This approach allows us to efficiently achieve physically relevant image structures from physically motivated parameters, without the long computational times required to realize them in GRMHD simulations. 

We take the following features as guiding assumptions for the design of our model:
\begin{enumerate}
    \item The emission is synchrotron radiation from a magnetized plasma.
    \item The accretion flow is relatively optically thin to $mm$ wavelengths \citep{Falcke_1998, galactic-center}.
    \item The emission region is compact and lies within a few gravitational radii of the black hole, with the bulk $mm$ wavelength emission often being produced in regions along the jet sheath and near the jet base  \citep{Dexter, Emami_2023}.
    \item The flow is taken to be axisymmetric, since it should reflect the time-averaged structure of the accretion flow.
    \item The emission is restricted to lie on a two oppositely oriented conical surfaces that are centered about the spin axis of the black hole.
\end{enumerate} 
The last assumption is motivated from observed locations of the peak mean emission within GRMHD simulations.
GRMHD simulations with thermal electron distribution functions in particular are typically seen to span a wide range of opening angles \citep[see for example Figures 3 \& 4 of][]{Emami_2023}.
The emission region in these simulations can range from either being flushed to the equatorial plane, a characteristic of low \rhigh MAD simulations, or significantly off the equatorial plane, as observed in high \rhigh SANE simulations \citepalias{EHTC_M87_V}.

\subsection{The Emission model}
\begin{figure}[htpb]
    \centering
    \includegraphics[width=0.45\textwidth]{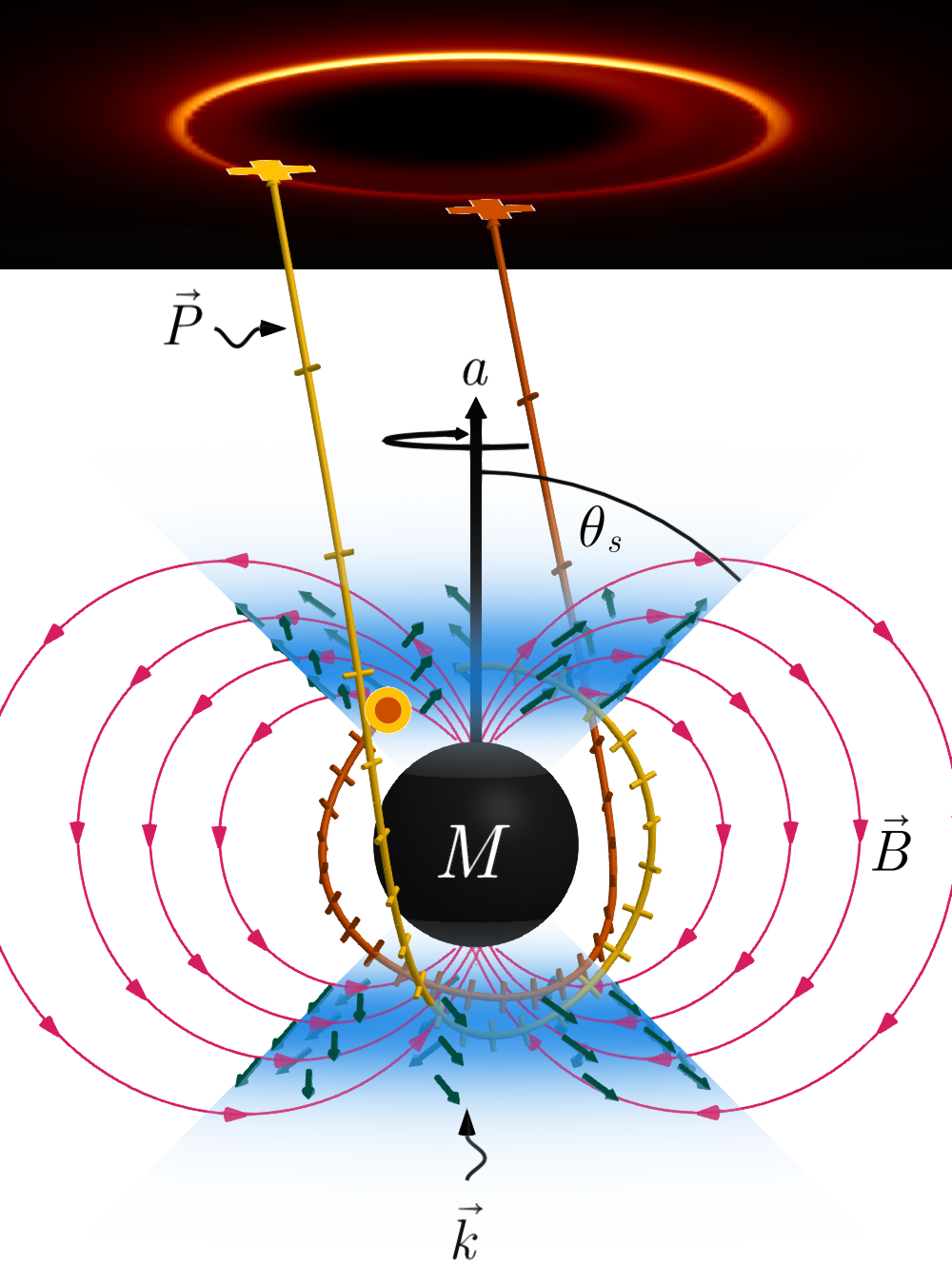}
    \caption{
        Schematic depiction of our dual cone emission model. 
        The full model consists of a Kerr black hole (black sphere) surrounded by synchrotron emitting plasma (green arrows) confined to a cone of opening angle $\theta_s$ and threaded by a magnetic field $\vec B$ (red field lines). 
        The magnetic field and plasma are taken to be axisymmetric about the equatorial plane. 
        The emission is concentrated at a characteristic radius, $R$, with the profile determined by a double power-law distribution with exponents $p_1$ and $p_2$ (translucent blue cone). 
        The polarization from the synchrotron radiation is parallel transported along null geodesics (yellow and orange ticked trajectories) to generate an image on the screen of an observer at radial infinity. 
        The image generated from the model is the sum of multiple sub-images corresponding to the different trajectories light can take from the source to the observer.}
    \label{fig:pedagogical}
\end{figure}
\autoref{fig:pedagogical} gives a cartoon schematic of the emission model.
Our model can be described as a compact, axisymmetric, linearly polarized emissivity, $j_\nu$, that is constrained to lie on a dual-cone with opening angle $\theta_s$. 
The emissivity is determined by the interactions of a magnetic field with a plasma flow around the black hole. 
The plasma flow is constrained to move along the surface of the cone, while the magnetic field can orient freely.

Following \citet{Palumbo_2022}, we impose a compactness constraint on the emission by choosing a radially dependent, double power law, isotropic emissivity profile given by,
\begin{align}
    \CMcal J(r, R, p_1, p_2)
        &=\frac{(r/R)^{p_1}}{1+(r/R)^{p_1+p_2}}\label{eqn:double_power_law},
\end{align}
where $r$ is the emission radius, $R$ is the characteristic radius of the profile, $p_1$ is the exponent of the inner power law, and $p_2$ is the exponent of the outer power law.
When given a magnetic field $B$ and plasma fluid velocity, $u$, this choice allows us to write our emissivity function as,
\begin{align}
    j_\nu(r, R, p_1, p_2,B, u)
        \propto \CMcal J(r, R, p_1, p_2 ) P_\nu(B, u).
\end{align}
where $P_\nu$ is a spatially isotropic function that gives the linear polarization intensity of synchrotron emission (we provide a simple power-law prescription for $P_\nu$ below). 
We ignore the missing proportionality constant and electron number density dependencies for this study since our analyses in \autoref{sec:image_domain} and \autoref{sec:visibility_domain} are independent of the overall flux of the image.

As with the emission models of \citet{Narayan_2021,Gelles, Palumbo_2022}, the calculation of $P_\nu$ is reliant on transformations to the frames of: the asymptotic observer (Boyer-Lindquist frame), a zero angular momentum observer (ZAMO), and the fluid frame.
Although our construction of these frames is qualitatively similar that of \citet{Narayan_2021} and \citet{Gelles},
we will summarize the procedure here since some difference are introduced by the non-equatorial emission geometry. 

The Kerr line element in Boyer-Lindquist coordinates is given by 
\citep{Chandrasekhar1984}
\begin{align}
    d s^2=-\frac{\Delta \Sigma}{\Xi} d t^2+\frac{\Sigma}{\Delta} d r^2+\Sigma d \theta^2+\frac{\Xi \sin ^2 \theta}{\Sigma}[d \phi-\omega d t]^2,
\end{align}
where,
\begin{align}
    \Delta(r)=r^2-2 M r+a^2, \quad \Sigma(r, \theta)=r^2+a^2 \cos ^2 \theta, \notag\\
    \omega=\frac{2 a M r}{\Xi}, \quad \Xi=\left(r^2+a^2\right)^2-\Delta a^2 \sin ^2 \theta .
\end{align}
The transformation for co-vector components from the global Boyer-Lindquist frame to the local ZAMO frame is then given by the tetrad, $e_{(m)}{}^\mu$, where we have used Greek indices to label components with respect to the global Boyer-Lindquist basis, and parenthesised Latin indices to label components with respect to the local ZAMO basis.
The components of the tetrads are \citep{Bardeen1972}\footnote{Our definition of the tetrad basis ordering matches the definitions in \citet{Narayan_2021,Palumbo_2022,Gelles}, but not the traditional basis ordering. 
    The negation of $e^{(3)}$ from its traditional definition ensures that the basis is always right handed.}
\begin{subequations}
\label{eqn:tetrads}
\begin{align}
    e_{(0)}
        & = \sqrt{\frac{\Xi_{\mathrm{s}}}{\Sigma_s\Delta_{\mathrm{s}}}}\left(\partial_t+\omega_{\mathrm{s}} \partial_\phi\right), \\
    e_{(1)}
        & = \sqrt{\frac{\Delta_{\mathrm{s}}}{\Sigma_s}} \partial_r, \\
    e_{(2)}
        & =\sqrt{\frac{\Sigma_s}{\Xi_{\mathrm{s}}}}\frac{1}{\sin\theta_s} \partial_\phi,\\ 
    e_{(3)}
        & =-\frac{1}{\sqrt{\Sigma_s}} \partial_\theta,
\end{align}
\end{subequations}
where $\Xi_s$, $\omega_s$, $\Sigma_s$ and $\Delta_s$ are the values of $\Xi$, $\omega$, $\Sigma$ and $\Delta$ evaluated at $r_s$ and $\theta_s$.
The ZAMO frame is labeled with directions $\{\hat x^0,\hat x^1,\hat x^2, \hat x^3\}$ which form a non-holonomic basis.
This choice of ZAMO basis orientation ensure that that $\hat x^1$ and $\hat x^2$ lie in the tangent space of the emission cone.

The fluid frame is defined from the fluid velocity vector,
\begin{align}
    u^{(m)}=\beta_v\langle0,\cos \chi,\sin \chi,0\rangle^{(m)},
\end{align}
which depends on an azimuthal angle $\chi$, that winds from the $\hat x^{(1)}$ direction to the $\hat x^{(2)}$ direction, and fluid speed $\beta_v$.
Since $\hat x^{(2)}||\hat \phi$, then the following cases define whether the fluid is in prograde or retrograde motion with respect to the hole,
\begin{align}
\begin{cases}
    \text{sign}(\chi)=\text{sign}(a)& \text{ Prograde}\\
    \text{sign}(\chi)\neq\text{sign}(a)& \text{ Retrograde}.
\end{cases}
\end{align}

Both, $\chi$ and $\beta_v$ are measured in the frame of the ZAMO. 
Our definition of the fluid velocity vector constrains it to lie in the tangent space of the dual-cone.
The fluid velocity vector can be used construct a fluid frame through Lorentz transformation from the ZAMO frame,
\begin{align}
    f^{(m')}=\Lambda^{(m')}{}_{(n)}f^{(n)},
\end{align}
where we have used the primed Latin indices to label the fluid frame components.

We assume that the polarization intensity scales as a power law,
\begin{align}
\label{eqn:polarization}
    P
        &\propto\delta^{3+\sigma}l_p 
        \lvert \mathbf B\rvert^{1+\sigma}\sin^{1+\sigma}\zeta.
\end{align}
This precise relationship can arise from power-law distributions for the electron number densities and energies, but it is approximately true under much more general cases \citep{Narayan_2021}.
The resulting polarization intensity is then determined by a spectral index ($\sigma$),
Doppler factor ($\delta$; which includes gravitational redshift and velocity effects), and the magnetic field in the fluid frame ($B^{(m')}$).
The remaining quantities are $l_p$, which is a proxy for the path length through a cone of width $H$ that depends on the emitting photon four momentum $p^{(m')}$ as,
\begin{align}
    l_p = \frac{p^{(0')}}{p^{(3')}} H,
\end{align} 
and $\zeta$, the angle associated with the cross product of $p^{(m')}$ and $B^{(m')}$ in the fluid frame,
\begin{align}
    \sin\zeta
        &=\frac{\lvert \mathbf{p}\times \mathbf{B}\rvert}{\lvert\mathbf p\rvert\lvert \mathbf B\rvert}\label{eqn:kcrossb}.
\end{align}
Here, the vector quantities $\mathbf{p}$ and $\mathbf{B}$ are the purely space-like components of $p^{(m')}$ and $B^{(m')}$.

We parameterized the direction of $B^{(m')}$ in terms of the fluid frame polar angle, $\iota$, and the fluid frame azimuthal angle $\eta$ as,
\begin{align}
 B^{(m')}
    &\propto\langle0,\sin(\iota)\cos(\eta),\sin(\iota)\sin(\eta),\cos(\iota)\rangle^{(m')}.
\end{align}

In general, the polarization direction is taken to be orthogonal to the magnetic field at the point of emission.
We will require that the radial and azimuthal components of the magnetic field in each cone have opposite sign as a proxy for the the lines being dragged by the plasma flow or black hole, and to ensure that the global magnetic field remains divergence free.
That is, for a given $\eta$ and $\iota$, then the magnetic fields can be written as:\footnote{We have relaxed the assumption of \citet{Narayan_2021} and \citet{Palumbo_2022} that relates $\eta$ and $\chi$, and have chosen instead to make them independent.}
\begin{align}
\left.\mathbf B\right\rvert_{\text{cone}_1}(\iota,\eta)
    &=
\left.\mathbf B\right\rvert_{\text{cone}_2}(-\iota,\eta).
\end{align}
The polarised emission is then raytraced, and parallel transported to a distant observer through conservation of the Walker-Penrose constant, \citep{Walker:1970un,Chandrasekhar1984},
\begin{align}
\kappa=(A-i B)(r-i a \cos \theta) \equiv \kappa_1+i \kappa_2, 
\end{align}
where,
\begin{align}
A
    &=(P^rp^t-p^rP^t)+a\sin^2\theta_s(P^\phi p^r - p^\phi P^r),\text{ and}\\ 
B
    &=\left[(r^2+a^2)(p^\phi P^\theta-p^\theta P^\phi)-a(p^tP^\theta-p^\theta P^t)\right]\sin\theta,
\end{align}
$p^\mu=(p^t,p^r,p^\theta,p^\phi)$ is the photon momentum in Boyer-Lindquist coordinates, 
and $P^\mu=(P^t,P^r,P^\theta,P^\phi)$ is the polarization vector.
Although our model naturally produces images with polarized emission structure, the rest of this work will be focused on studies of the total intensity.

\subsubsection{Raytracing}
The construction of our model allows us to produce images using analytic solutions of the Kerr geodesic equations.
This fact makes evaluation of our model extremely efficient, a necessary condition for performing posterior explorations of fits to realistic data, which will be discussed in \autoref{sec:visibility_domain}.
Our ray-tracing procedure is similar to that of \cite{Gelles}, with the exception that the presence of non-equatorial emitters requires; (i) the inclusion of vortical trajectories to the geodesic solution space, and (ii) a new definition of sub-image indexing that is a generalization from ones used by previous authors \citep[e.g.,][]{Bao,MJohnsonRing,patoka}. 
We will briefly outline how we use the solutions to the geodesic equations to raytrace our model; \autoref{sec:image_index} provides details on our image indexing definition.

The solutions to the geodesic equations can be found by reduction to quadratures through the Hamilton-Jacobi approach \citep{Carter,Dexter2009,GrallaLupsasca,Himwich}. 
The four-momenta, $p_\mu=g_{\mu\nu}\frac{dx^\mu}{d\tau'}$, of the photons for an affine parameter $\tau'$ are then written as,
\begin{align}
    p_\mu dx^\mu
        &=-dt \pm_r\frac{\sqrt{\CMcal R(r)}}{\Delta(r)}dr\pm_\theta\sqrt{\Theta(\theta)}d\theta +\lambda d\phi,
\end{align}
where,
\begin{align}
    \CMcal R (r)
        &= \left(r^2+a^2-a \lambda\right)^2-\Delta\left[\eta+(a-\lambda)^2\right]\label{eqn:radial_potential}
\end{align}
is the radial potential,
\begin{align}
    \Theta(\theta)
        &=\eta+a^2 \cos ^2 \theta-\lambda^2 \cot ^2 \theta\label{eqn:theta_potential},
\end{align}
is the inclination or $\theta$-potential, $\lambda$ is the energy reduced azimuthal angular momentum, and $\eta$ is the energy reduced Carter integral.

When expressed in this form, it is typical to parameterize the solutions to the geodesic equations in terms of the Mino-time, $\tau$ \citep{Mino_2003}:
\begin{align}
   \Sigma(r,\theta) d\tau=d\tau'.
\end{align}
The Mino-time can then be expressed as integrals in terms of either $r$ or $\theta$, which both have solutions in terms of elliptic integrals \citep[for a review of the solutions to the Kerr geodesic equations, see][]{Rauch_Blandford,GrallaLupsasca}:
\begin{align}
    \Delta\tau
        =\fint_{r_s}^{r_o}\frac{d r}{\sqrt{\CMcal R(r)}} 
        &=\fint_{\theta_s}^{\theta_{\rm o}}\frac{d\theta}{\sqrt{\Theta(\theta)}} .
\end{align}
The elliptic integrals can be inverted to allow their amplitudes to be expressed in terms of Jacobi elliptic functions.
It is therefore possible to write expressions which relate the emission radius of a null-geodesic in terms of emission inclination, $\theta_s$, observation inclination, $\theta_{\rm o}$, observer screen coordinates, $(\alpha,\beta)$, and the order of sub-image, $n$,
\begin{align}
    r_s=r(\theta_s, \theta_{\rm o}, \alpha, \beta, n),
\end{align}
without the need for a numerical raytracing algorithm that solve the second order geodesic equations \citep[e.g.][]{Moscibrodzka, Raptor}.

We have taken these properties of the Kerr geodesics into consideration through the implementation of a fast, GPU-compatible, ray-tracing code for geometries embedded in Kerr spacetimes \texttt{Krang.jl}\footnote{\href{https://github.com/dominic-chang/Krang.jl}{\textcolor{blue}{\underline{https://github.com/dominic-chang/Krang.jl}}}}.
We additionally present a GPU-compatible implementation of multiple algorithms for calculating elliptic integrals and Jacobi elliptic functions \citep[]{Fukushima, Carlson_1995} through the code \texttt{JacobiElliptic.jl}
\footnote{
\href{https://github.com/dominic-chang/JacobiElliptic.jl}{\textcolor{blue}{\underline{https://github.com/dominic-chang/JacobiElliptic.jl}}}.
Our convention here disagrees with \citet{NIST:DLMF} but is in agreement with \citet{GrallaLupsasca}. 
The two are related by the relationships:
$F(\phi, \sqrt{k}) =F(\phi \mid k)$,
$E(\phi, \sqrt{k}) =E(\phi \mid k)$  and $\Pi(\phi, n, \sqrt{k}) =\Pi(n ; \phi \mid k)$.
}.
Both codes are compatible with CUDA-compatible GPUs and the GPUs on the Apple M-series chips.

\section{Image domain comparisons with GRMHD}
\label{sec:image_domain}

We test the feasibility of using the dual-cone model to represent the horizon scale structure of AGN cores by studying its ability to reproduce images of time-averaged GRMHD models of \sgra and \m87.
These include GRMHD simulations of MAD (magnetically arrested disk) and SANE (standard and normal) accretion flows which are described by the strength of their magnetic flux across the midplane.
MAD models typically feature high magnetic pressure, with ordered field lines, and greater jet power, while SANE models feature higher fluid pressure, and more turbulent flows. 
The MAD accretion flows are of particular interest since they exhibit more jet power than their SANE counterparts, and are largely more consistent with EHT data \citepalias[see analyses of][]{EHTC_M87_VI, EHTC_SgrA_V}.
For our study, we use time-averaged \texttt{iHARM} and \texttt{KHARMA} GRMHD models of MAD and SANE accretion flows from the Illinois simulation library, \citep{iharm,EHTC_M87_V,EHTC_SgrA_V}, whose emissivities are extracted and imaged with the \texttt{ipole} GRRT (General Relativistic Radiative Transfer) code in the \texttt{PATOKA} pipeline, \citep{ipole,patoka}.
The emission from these models is dependent on a simple prescription for the ion-to-electron temperature ratio of the fluid:
\begin{align}
    \frac{T_i}{T_e}
        &=\text{\rhigh}\frac{b^2}{1+b^2}+R_{\text{low}}\frac{1}{1+b^2}.
\end{align}
Here, $b=\beta/\beta_{\text{crit}}$ where $\beta=P_{\text{gas}}/P_{\text{mag}}$ gives the ratio of the gas-to-magnetic pressure. The two tuning parameters \rhigh and $R_{\text{low}}$ tend to describe the temperature ratios in the (weakly magnetized) disk and the (strongly magnetized) jet, respectively.


The analyses in this section were performed using a template matching algorithm implemented in \texttt{VIDA.jl} \citep{VIDA} to generate image domain representations of GRMHD from a set of 3 different model classes.
We use \texttt{BlackBoxOptim.jl}, a \texttt{julia} implementation of a differential evolution algorithm, generate images from our model that are `represent' those of GRMHD by minimizing the loss function between images.
We take the log of the magnitude of the normalized cross correlation (NxCorr) between two $N\times N$ images $A$ and $B$ as our loss function.
The $\log(\lvert\text{NxCorr}\rvert)$ is given by;
\begin{align}
    \text{NxCorr}&=\left\lvert\frac{
    \sum_{i}
    (A_i-\mu_A)(B_i-\mu_B)}{
    \sqrt{\sum_{i}(A_i-\mu_A)^2}\sqrt{\sum_{j}(B_i-\mu_B)^2}
    }\right\rvert,
\end{align}
where $\mu_A$ and $\mu_B$ are the average intensity of the two images, and $i,j\in\{1,\dots,N\}$ index their pixels.
We will refer to the models that generate these images as `representative models'.

\begin{figure}[htpb]
    \centering
    \includegraphics[width=0.47\textwidth]{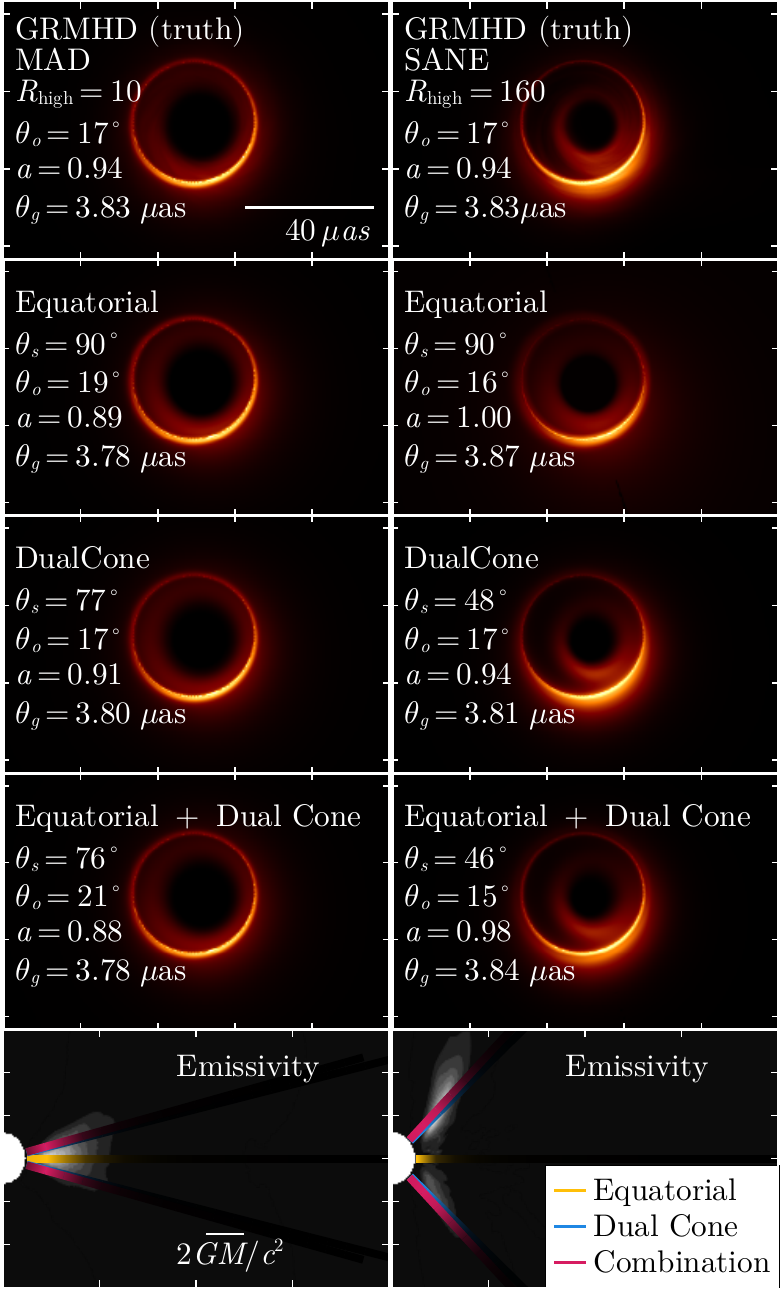}
    \caption{
        Comparison of time-averaged MAD (left) and SANE (right) GRMHD simulations of \m87 (top row) with the best-fitting equatorial (2nd row), dual-cone (3rd row), and combination (equatorial + dual cone) models (4th row). These two simulations are representative of the minimal and maximal emission scale heights that are typically seen in GRMHD simulations of \m87. The best-fitting models are determined by maximizing the normalized cross-correlation in the image domain. The text in each panel gives the opening angle of the cone ($\theta_s$), the viewing inclination ($\theta_{\rm o}$), the black hole spin ($a$), and the mass-to-distance ratio ($\theta_g$).
        The bottom row compares the fitted emissivity functions of our models (colored lines) to the true emissivity of the GRMHD (grayscale).
        The predicted emissivities are consistent with those of the GRMHD, as is evident by their overlap, which illustrates our model's ability to perform photogrammetry in black hole space times.
    }
    \label{fig:emissivity}
\end{figure}

\subsection{Image Domain Fit: Simulations of \m87}
We compare the performance of three different classes of analytic models with different emission geometries.
These are: (i) an equatorial model, similar to the one described in \citet{Palumbo_2022}, (ii) our dual-cone model described in \autoref{sec:theory}, and (iii) a `combination' model whose emission is the sum of the two.
These models are used to generate images that are similar to those that are produced by time averaging a $a=0.94$, prograde, \rhigh $=10$, MAD GRMHD simulation and a $a=0.94$ spin, prograde, \rhigh $=160$, SANE simulation.
Both simulations are  taken to have mass-to-distance ratios of $3.83\mu$as, matching EHT measurements of \m87 \citepalias{EHTC_M87_V}.
An equivalent fit to an $a=0.5$ \rhigh $=20$ MAD and an $a=0.5$ \rhigh $=160$ SANE are shown in the \autoref{fig:m87-nxcorr-summary} in \autoref{sec:image-domain-appendix}.
The parameter search range used to find the representative images of our models is detailed in \autoref{tab:best_nxcorr}.

\autoref{fig:emissivity} shows a summary of images of our representative models. 
These are all compared to the GRMHD model they were fitted to in the first row.
The general effect of \rhigh is to tune the relative emission contribution from the jet and the disk. 
Although trends with respect to \rhigh differ in MADs and SANEs, due to their different magnetic field strengths in the inner accretion flow \citep[compare, for example, the emissivities between models in][]{EHTC_M87_V, Emami_2023}, the simulations in the top row of \autoref{fig:emissivity} are typical of the extreme cases of emission scale heights seen in GRMHD.
In particular, low $R_{\text{high}}$ MAD simulations tend to have emission geometries with lower scale heights (more disk dominated) and high $R_{\text{high}}$ SANE simulations tend to have emission geometries with high scale heights (more jet dominated).

The increased emission contribution from the forward jet in the images of the SANE model of \autoref{fig:emissivity} manifests as a smaller secondary ring feature. 
The morphology of this feature is difficult for simple equatorial models to produce, as is evident when comparing the true images to the images of the equatorial model (second row in \autoref{fig:emissivity}) to the true images of the GRMHD.
The existence of such features is, however, captured by models capable of producing jet-like emission, as is apparent
in the images of the geometric models with off-equatorial components (the dual-cone and combination models of the third and fourth rows).

Of interest are the inferred bulk parameters of the accretion flow and spacetime from our models.
We note, for example, that the fitted mass-to-distance ratio, spin and observer inclinations of all our models are correct to within $3\%$ , $5\%$ and $10\%$ respectively.
The left and right panels in the last row of \autoref{fig:emissivity} shows the emission geometries of each of the representative models overlayed on the true emissivity contours of the MAD and SANE GRMHD simulations respectively.
All representative models capture the compactness of the peak emission, but only the dual-cone and combination models are able to capture the scale height of the true emission geometries.
Thus the dual-cone models appear capable of performing accurate photogrammetry of the emission geometry from images of GRMHD.

\begin{figure*}[htpb]
    \centering
    \includegraphics[width=\textwidth]{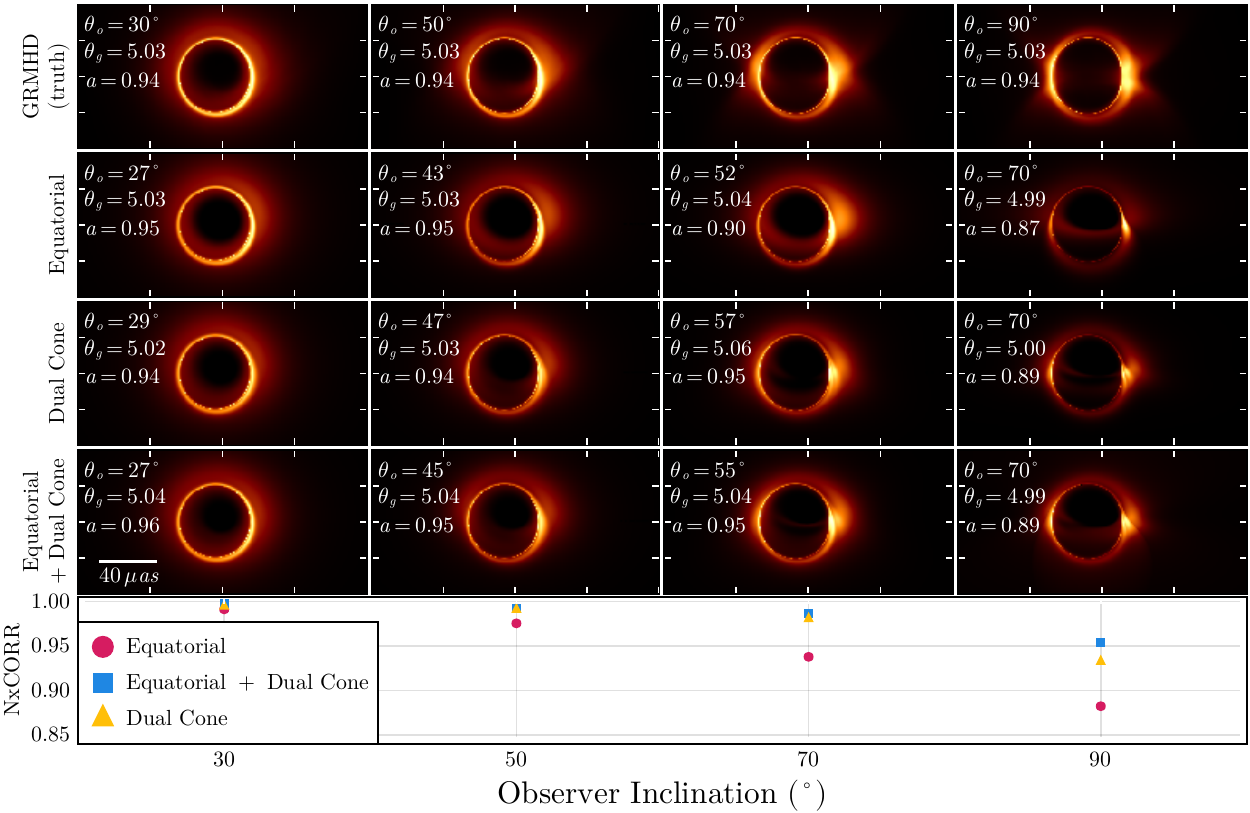}
    \caption{
    Best-fitting equatorial, dual cone, and combination (equatorial + dual cone) models to a time-averaged retrograde \rhigh$=160$ MAD GRMHD simulation of \sgra viewed at inclinations of $30^\circ$, $50^\circ$, $70^\circ$, and $90^\circ$ (from left to right). The NxCORR score for all models is above 0.85 and increases as the black hole is viewed more face-on. In addition, the parameters of the best-fitting models are within a few percent of the true parameters for both mass and spin at low-to-modest inclinations. 
    As expected, the dual cone model (13 parameters) outperforms the equatorial model (12 parameters) despite having only a single additional parameter. In contrast, the combination model (23 parameters) provides only marginal additional improvement despite nearly doubling the number of parameters. \autoref{tab:best_nxcorr} lists the parameters for each model.
    }
    \label{fig:nxcorr_summary}
\end{figure*}

\subsection{Image Domain Fit: Simulations of \sgra}
In models of \m87, such as those in \autoref{fig:emissivity}, the low observer inclination allows for two-dimensional emissivities to adequately represent the screen projection of the three-dimensional structures seen in GRMHD.
Images of simulations viewed at higher inclinations, in contrast, may be more sensitive to edge-on effects that are not captured by our model.
We evaluate the performance of our models at fitting images taken at varying observer inclinations by fitting images of $R_{\text{high}}=160$ GRMHD simulations of \sgra from the Illinois simulation registry.
These simulations were raytraced to produce images at inclinations of $30^\circ$, $50^\circ$, $70^\circ$ and $90^\circ$, (see \autoref{fig:nxcorr_summary}). 

All models are again able to recover the true $\theta_g$ values of the source within good accuracy, fitting values within $1\%$ of the truth, but tend to a have a bias towards inferring lower masses as the tilt of the source increases. 
The models tend to be less accurate on spin inferences at high inclinations, but perform well at low and moderate inclinations of $30^\circ$, $50^\circ$ and $70^\circ$, capturing  the true spin to within $5\%$.
The decreased accuracy of our model at recovering the true black hole parameters at moderate and high inclinations is likely a result it providing worse image representations of structures seen at higher inclinations. 
Our the conical emission geometry is innately worse at predicting edge-on image structures corresponding to thick disks; as expected, there is an inverse relationship between NxCORR performance and source inclination.
\begin{figure*}[htpb]
    \centering
    \includegraphics[width=\textwidth]{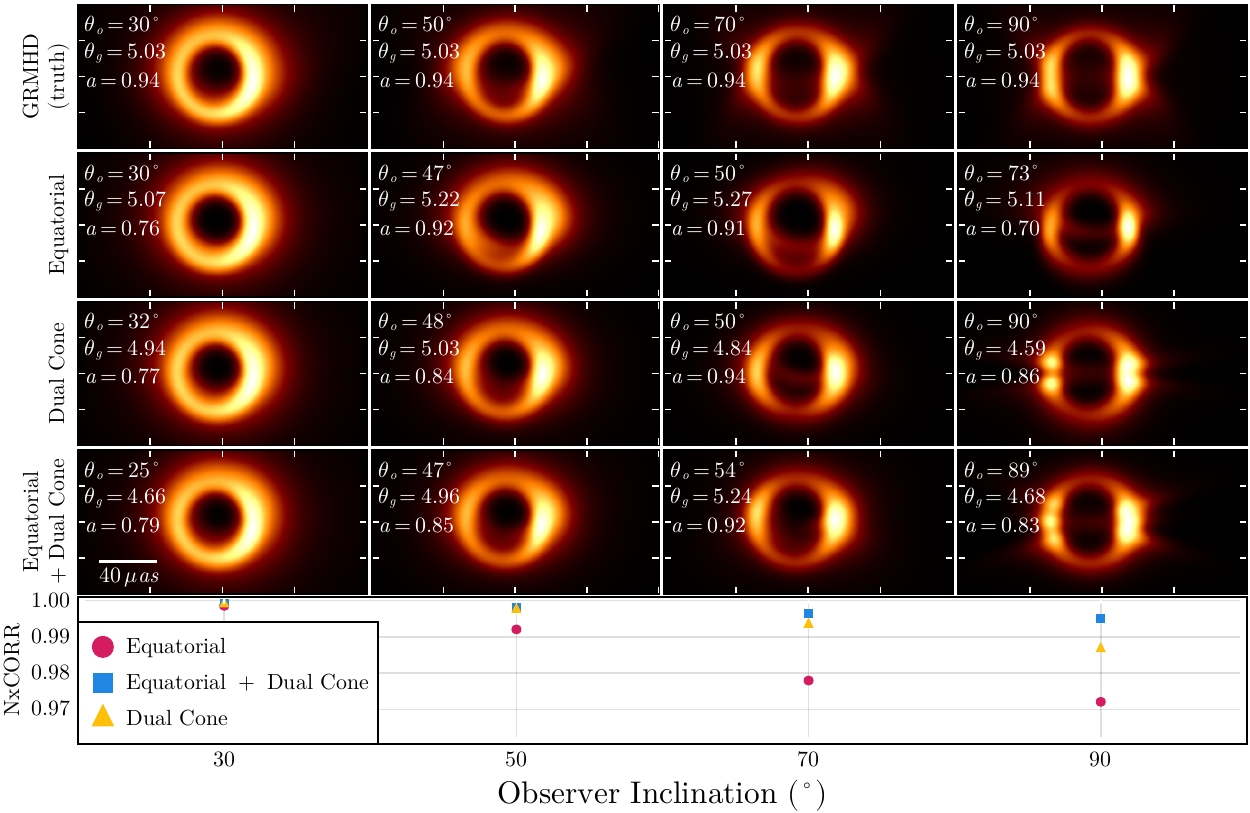}
    \caption{
    Same as \autoref{fig:nxcorr_summary}, but when the ground truth image and model image are both blurred by a $10\,\mu{\rm as}$ FWHM Gaussian kernel before comparison. The resulting NxCORR values are higher than for the unblurred case, with excellent fidelity for all model types. However, despite the improved image fidelity at this resolution, the best-fitting parameters have significantly larger discrepancies than for the unblurred fits, showing the strong degeneracies that are present when restricted to the current EHT resolution, especially because the photon ring cannot be distinguished from the direct emission.     
    }
    \label{fig:nxcorr_blur_summary}
\end{figure*}

We also study the effects of limited resolution on the inference capabilities of our model. 
We perform this study by executing the same comparison as done in \autoref{fig:nxcorr_summary}, but after convolving both the GRMHD images and the model images with a Gaussian beam of $10 \mu\text{as}$ FWHM during the fitting process.
This blurring kernel approximates the results of imaging EHT sources.
A summary of the results is show in \autoref{fig:nxcorr_blur_summary}.

The inverse relationship between the model's NxCORR scores and inclination still remain, although the overall performance of the model has improved.
This indicates that our model accurately reproduces the structures in the images of time-averaged GRMHD that can be accessed with the resolution of current instruments. 
However, while the effect of blurring improves the NxCORR in the best fits, the parameter inference is significant worse, especially for $\theta_g$ and spin.
This effect indicates that our model's ability to infer mass and spin is likely sensitive to fine scale image structures that are no longer present in the image after blurring. 
One possibility is the tendency for spin to cause small asymmetries in the black hole shadow \citep[e.g.,][]{Takahashi} and to introduce shifts in the photon ring with respect to the direct emission \citep[e.g.,][]{Papoutsis}.
These effects are diminished once the GRMHD models are blurred to EHT resolutions.
Features that largely survive blurring include the brightness asymmetry and apparent shadow diameter.
The shadow diameter is mostly consistent at all inclinations, which suggests that the model's ability to infer the true inclination is likely dependent on the morphology of the brightness distribution in the image. 

\begin{figure*}
    \centering
    \includegraphics[width=\textwidth]{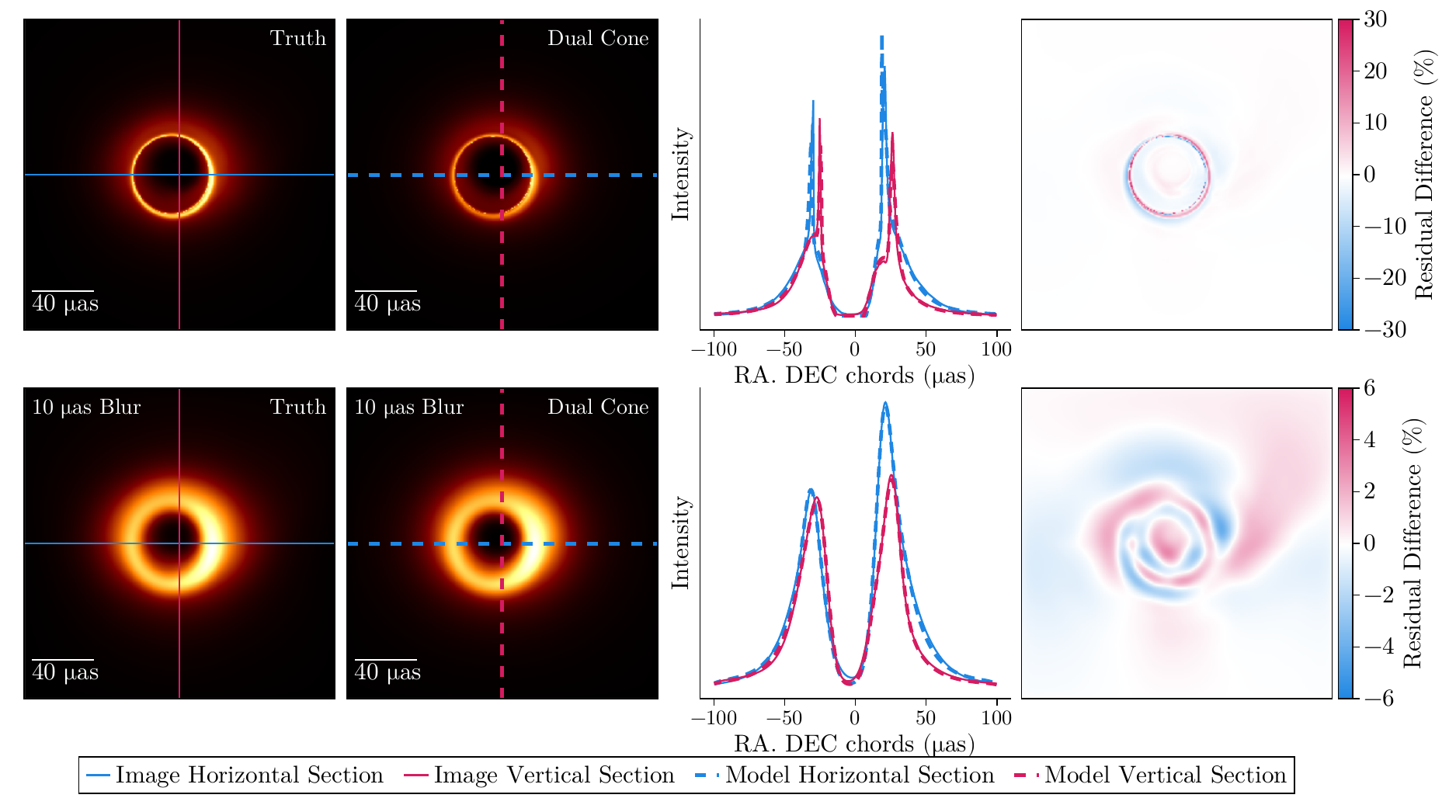}
    \caption{Comparison of a time-averaged GRMHD image with the best-fit Dual Cone model, both at native resolution (top) and after being blurred by a $10\mu$as FWHM Gaussian beam (bottom). The simulation corresponds to a retrograde, \rhigh$=160$, MAD simulation of \sgra. Columns show the GRMHD image (left), the best-fit model (left-center), horizontal and vertical intensity cross sections (right-center), and residual images as a fraction of peak intensity (right).
    }
    \label{fig:image-residual}
\end{figure*}

We emphasize that these simple dual cone emission models provide accurate representations of the image domain morphology of time-averaged GRMHD, even when blurred to EHT resolution.
\autoref{fig:image-residual} shows two representative examples of a best fit NxCORR dual-cone model along with its truth image at native resolution and after being blurred by a 10 FWHM Gaussian beam.
The intensity profiles of the dual cone model are strikingly similar to the truth, especially along the photon ring region.
This is an interesting result since the dual cone model is taken to be completely optically thin, and thus would be expected to produce generically brighter photon rings than what would be seen from a more accurate model with optical depth effects, such as GRMHD.
This results indicates that there is likely some degeneracy between optical depth effects the emission physics effects of our models. 
For example, our optically thin model may exploit the anisotropy of synchrotron emission to replicate the impact of optical depth on tuning the relative flux between the photon rings of models with different emission physics.
It is then possible for anisotropy effects to cause the $n=1$ ring to appear dimmer than would be expected by partitioning the flux of emitters away from the $n=1$ direction \citep[see][ for an example of how emission anisotropy can affect the total flux seen by an observer at ininity from equatorial emitters, and Figure~3 of \citet{Bardeen_Cunningham} for an example of how flux in the $n=0$ and $n=1$ images are partitioned by emission direction]{Gates_2020}.

\section{Bayesian inference in the visibility domain}
\label{sec:visibility_domain}
VLBI observations of a source give rise to sparse data products in the Fourier domain of an image. 
Thus, interpreting VLBI data is reliant on a source model, regardless of whether the model's purpose is imaging or parameter extraction. The EHTC has used many approaches to produce images from VLBI observations \citepalias[see, e.g.,][]{EHTC_M87_IV,EHTC_SgrA_IV}. 
While flexible, these algorithms are only able to constrain the image domain structure and not the physical source parameters (e.g., black hole mass and spin) directly. Therefore, converting simple geometric properties of the data/images, such as ring size, shape, and width, to intrinsic properties of the physical system requires calibration. 

In \citetalias{EHTC_M87_VI}, \citetalias{EHTC_SgrA_IV}, and \citetalias{EHTC_SgrA_VI} a calibration procedure called $\alpha$-calibration was used to move from the ring size to the mass of the central black hole. 
By creating a large number of synthetic data sets of GRMHD snapshots, each imaging or geometric modeling pipeline was able to relate the reconstructed diameter $\hat{d}$ to the intrinsic gravitational radius of the system $\theta_g$
using the equation
\begin{equation}
    d = \alpha\theta_g,
\end{equation}
recovering a GRMHD-based calibration for $\alpha$, and relating the measured ring size of M87*/Sgr~A* to the mass of the central black hole. 

However, this analysis is not ideal. In particular, it is likely that significant information from the black hole's influence on its image is not incorporated into the calibration procedure since it only uses the single parameter for the ring diameter. Any imposed constraints from additional relativistic features are not translated through the $\alpha$-calibration into the GRMHD solution space.

An ideal analysis would involve directly fitting dynamic accretion models to data, for example, directly fitting GRMHD and all of its accretion and emission parameters to the data. 
Unfortunately, the ability to conduct such an analysis with GRMHD was and is likely infeasible due to the computational complexity of directly sampling the solution space.

Following \citet{Palumbo_2022}, we avoid the two-step procedure of imaging followed by $\alpha$-calibration by using our dual cone model as a method to infer physical parameters directly from VLBI data.
We first perform a Bayesian inference study of the dual cone model, using 2 different data sets.
The first is generated from a synthetic observation using an instantiation of the Dual Cone model. 
For this test, we use the model that has the best NxCORR when fit to the $R_{\text{high}}=160$ SANE GRMHD of \autoref{fig:emissivity} whose flux been normalized to that of the GRMHD simulation that it was fitted from.
A study on this data is important for understanding degeneracies that may exist when performing inference on realistic data.
The second data set comes from a synthetic observation of the $R_{\text{high}}=160$ SANE GRMHD simulation that was used to create the previously mentioned model instance. 
This data set allows us to compare the visibility-domain fits to the image-domain NxCORR analysis to understand how the limited coverage of the EHT affects inferred parameters.
Finally, we use data from the public release of the EHT April~6 2017 observations of \m87 to perform a Bayesian parameter inference study on \m87. 
This sequence of tests allows us to assess the successes and shortcomings of our model under a variety of increasingly challenging (and realistic) circumstances, with increasing degrees of known and unknown model errors. 
We report corner plots of the parameter posteriors for all fits, and report the highest probability density interval (HPDI) of each parameter which is the smallest interval of a parameter that contains 95\% of the posterior mass.

\subsection{Data products}

The ideal visibilities of a source are related to image intensities, $P(x,y)$, through a 
Fourier transform,
\begin{align}
    \mathcal{V}_{ij}(u_{ij}, v_{ij})
        &=\iint \mathrm{e}^{-2 \pi i(u_{ij} x+v_{ij} y)} P(x, y)\; d x\,d y,
\end{align}
where $i,j$ indexes the two sites that make up the baseline at the point $(u_{ij}, v_{ij})$ in visibility space, and $(x, y)$ are the horizontal and vertical screen coordinates.
However, the ideal visibilities differ from the observed visibilities due to the additional effects of noise and site-specific corruptions (called ``gains'').
In particular, the observed visibilities, $V_{ij}(u_{ij},v_{ij})$ are related to the ideal visibilities through the relationship, 
\begin{align}
    V_{i j}
        &=g_i g_j^* \mathcal{V}_{i j}+\epsilon_{i j}=\left|V_{i j}\right| e^{i \phi_{i j}},
\end{align}
where $g_i$ are the site specific gain corruptions \citep{TMS}.
The gain corruptions, $g_i$, are effectively modeled as complex numbers that act independently at each site.\footnote{Interestingly, site-specific corruptions allow for an interpretation of the observed visibilities as a discontinuous field with a local $\text{GL}(1,\mathbb{C})$ symmetry \citep{Samuel}.} 
One can then define gain invariant quantities from closed triangles and quadrangles of visibilities.
These gain invariant quantities are the closure phases, 
\begin{align}
    \psi_{i j k}
        &=\arg \left(V_{i j} V_{j k} V_{k i}\right)=\phi_{i j}+\phi_{j k}+\phi_{k i},
\end{align}
and log closure amplitudes,
\begin{align}
    c_{i j k l}=\ln \left|\frac{V_{i j} V_{k l}}{V_{i k} V_{j l}}\right|=a_{i j}+a_{k l}-a_{i k}-a_{j l}.
\end{align}
The $\phi_{ij}$ and $a_{ij}$ are the visibility phases and log visibility amplitudes, respectively, and are related to the observed complex visibilities by,
\begin{align}
    V_{ij}
        &=\exp(a_{ij} + i\phi_{ij})
\end{align}
These closure quantities have been used by many authors in past analyses of EHT data, \citep[e.g.,][]{chael,EHTC_M87_III,EHTC_M87_IV,EHTC_M87_VI,Narayan_2021, Lockhart_Gralla, Medeiros}.

We generate the visibilities for the data sets used in the self-fit and GRMHD fits with the \texttt{observe\_same} functionality of \texttt{eht-imaging} to mimic EHT-like coverage from the April~6th observations of \m87.
All data used has been averaged over each scan using the \texttt{scan\_average} functionality of \texttt{Pyehtim}, a \texttt{julia} wrapper of the python library \texttt{eht-imaging}.
Since the large-scale structure around \m87, which we do not model, is over resolved by the EHT array, we flag visibilities with $u-v$ distances shorter than $0.1\,{\rm G}\lambda$. 
All visibility data products of the dual-cone model in this section are generated from $180{\times}180$ pixel images on a $120\,\mu\text{as} \times 120\,\mu\text{as}$ field of view.

\subsection{Log-Likelihoods}

The likelihoods for a set of non-redundant closure quantities can be constructed under the assumption that they are characterized by a multivariate Gaussian distribution  \citep{Blackburn}. 
The closure amplitude likelihoods, in particular, can be written as,
\begin{align}
    \mathcal{L}( \mathbf{c}\mid  \mathbf{\hat c})
        &\propto\exp\left[-\frac{1}{2}\mathbf{\tilde c}^{\intercal}\mathbf{\Sigma}^{-1}_{\mathbf{c}}\mathbf{\tilde c}\right],
\end{align}
where $\mathbf{\tilde c} = \mathbf{c}-\mathbf{\hat c}$ are a vector of residual closure amplitudes between the measured $\mathbf{c}$ and the model hypothesis $\mathbf{\hat c}$ and $\Sigma_\mathbf{c}$ is the covariance matrix of $\mathbf{c}$.
The likelihood for closure phases can be constructed similarly from their complex exponentials.
We define,
\begin{align}
    \mathbf e(\mathbf\Psi)
        &=\exp(i\mathbf{\Psi}),
\end{align}
to be a vector of complex exponentials formed from its element-wise action on vector of closure phases $\Psi$.
The likelihood is then,
\begin{align}
    \mathcal{L}(\mathbf\Psi\mid\mathbf{\tilde\Psi})
        &\propto\exp\left[
            -\frac{1}{2}
             \mathbf{\tilde e^\dagger}\mathbf{\Sigma}^{-1}_{\mathbf{e}}\mathbf{\tilde e}
        \right],
\end{align}
where we have chosen this form to account for the phase wrapping in $\psi$.
We do not include the data-dependent normalization coefficients of the likelihood functions since they contribute an overall constant to our posterior densities and, thus, do not influence inference.
\subsection{Priors}
\begin{table*}[htbp]
\centering
\begin{center}
\setlength\extrarowheight{3pt}
\setlength{\tabcolsep}{3pt}
\begin{tabular}{c|lcc} \hline \hline
\textbf{Parameter} & \textbf{Description} & \textbf{Units} & \textbf{Prior} \\ \hline
$\theta_g$ & Mass-to-distance ratio & $\mu\text{as}$ & $\CMcal{U}(1.5, 8)$ \\
$a$ & Black hole dimensionless spin & \dots & $\CMcal{U}(-1, 0)$ \\
$\theta_{\rm o}$ & Observer inclination& rad & $\CMcal{U}\left(1, \frac{40}{180}\pi\right)$ \\
$\theta_s$ & Cone opening angle& rad& $\CMcal{U}\left(\frac{40}{180}\pi, \frac{\pi}{2}\right)$ \\
p.a. & Position angle of projected spin axis& rad & $\CMcal{U}(-\pi, 0)$ \\
$R$ & Characteristic radius of number density function& $\frac{GM}{c^2}$ & $\CMcal{U}(1, 18)$ \\
$p_1$ & Inner exponent of the number density function& \dots & $\CMcal{U}(0.1, 10)$ \\
$p_2$ & Outer exponent of the number density function& \dots & $\CMcal{U}(1, 10)$ \\
$\chi$ & Fluid velocity azimuthal angle in ZAMO frame & rad & $\CMcal{U}(-\pi, \pi)$ \\
$\iota$ & Magnetic field orthogonality angle in fluid frame & rad & $\CMcal{U}(0, \frac{\pi}{2})$ \\
$\beta$ & Fluid speed in ZAMO frame & $c$ & $\CMcal{U}(0, 0.9)$ \\
$\sigma$ & Spectral index of Emission & \dots & $\CMcal{U}(-1, 3)$ \\
$\eta$ & Magnetic field tangential angle in fluid frame & rad & $\CMcal{U}(-\pi,\pi)$ \\
\hline
\end{tabular}
\end{center}
\caption{Model parameters and prior ranges used in the Bayesian analyses with the Dual Cone model (\autoref{sec:visibility_domain}).}
\label{tab:bayesian_prior}
\end{table*}
We generally take broad priors on our model to reflect our ignorance of the source.
Mass estimates of \m87 from stellar dynamics and gas dynamics methods are typically dependent on the distance from the Earth to \m87. 
The quantity constrained by each method is the mass-to-distance ratio, $\theta_g$. 
We, therefore, choose to report the measured $\theta_g$ of the source instead of its mass. 
The literature features widely discrepant measurements for $\theta_g$.
These measurements range from their smallest value of $\sim2\,\mu \text{as}$, from the gas dynamics methods to their largest of  $\sim6\,\mu\text{as}$, from the largest stellar dynamics measurements.
We therefore adopt wide priors on $\theta_g$ to encompass all of these, ranging from $1\,\mu\text{as}$ to $8\,\mu\text{as}$.

Priors on the spin of \m87 can be taken from properties of the large-scale jet and horizon scale images of \m87. 
We will assume that the orientation of the spin axis is related to the inclination and position angle of the large-scale jet.
Studies of the jet dynamics of \m87 suggest an inclination of $\sim17^\circ$ from line of sight, while images of the large scale jet imply a position angle of $288^\circ$ from north \citep[e.g.,][]{MLWH,Walker}. 
Horizon-scale images of \m87 show a peak brightness in the south, indicating that the spin of \m87 points away from our line of sight \citepalias{EHTC_M87_V}.
We incorporate these observations into our prior by fixing the observer's inclination to lie strictly within a range of $\sim 20^\circ$ around the observed large-scale jet inclination, $\theta_o\in[1^\circ,40^\circ]$, and fixing the spin to take on only negative values with the full range of magnitudes, $a\in[-1,0)$. 
Since the large scale jet observed pointing in an westerly direction on sky, the position angle shift is restricted to lie a range of, $p.a.\in[-\pi,0]$.
Description of all the priors used in the analyses of this section is given in \autoref{tab:bayesian_prior}.

\subsection{Sampling}
Our Bayesian analysis was performed within \texttt{Comrade.jl}, a VLBI statistical inference framework \citep{Tiede2022}.
Posterior sampling was performed with \texttt{Pigeons.jl} \citep{surjanovic2023pigeonsjl}, a \texttt{Julia} \citep{julia} implementation of a Non-reversible parallel tempering algorithm \citep{NRPT}.
We use the default slice sampler for the kernel of the exploration step \citep{Neal}.

\subsection{Visibility Domain Fit: Dual-Cone Model}
\label{sec:visibility-domain-self-fit}

\begin{figure}[htpb]
    \includegraphics[width=0.475\textwidth]{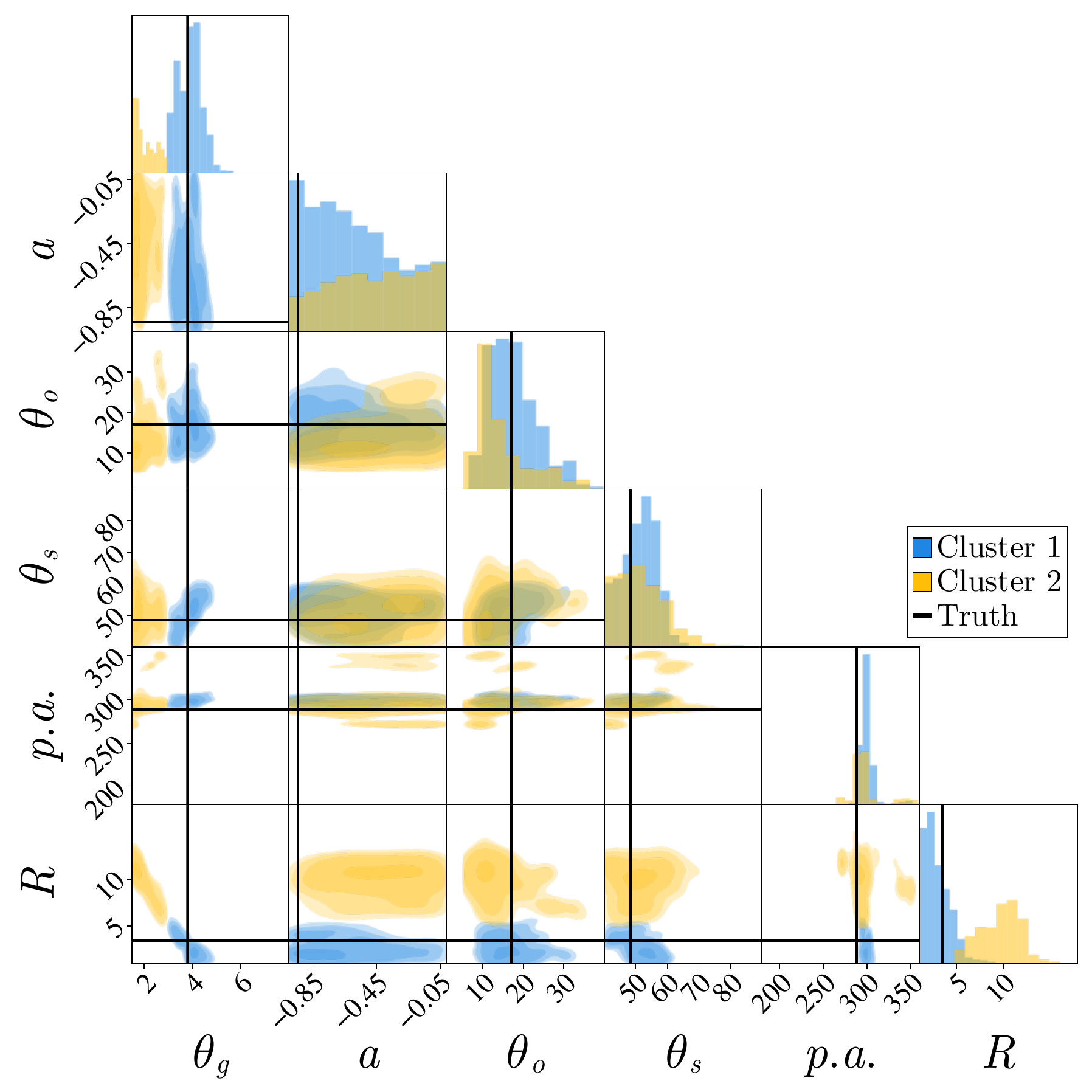}
    \caption{
        Partial corner plot showing the posterior for a subset of the parameter space of a dual cone fit to itself using EHT 2017 coverage. 
        The shown parameters are the mass-to-distance ratio ($\theta_g$), the black hole spin ($a$), the observer inclination ($\theta_{\rm o}$), the jet opening angle ($\theta_s$), the projected position angle of the spin axis on the observer's screen ($p.a.$), and the characteristic radius of the emissivity function ($R$).
        The true values of the model used to generate the data are shown with the heavy black lines. 
        The posterior features two distinct clusters, depicted in blue and yellow, which we isolate using the k-means clustering algorithm.
        \autoref{fig:visibility-self-fit-full} shows the full corner plot for this fit.
    }
    \label{fig:visibility-self-fit}
\end{figure}

This section features the results of a self-fitting exercise to synthetic data generated from our model. 
The synthetic data is generated using the coverage and the measured thermal noise of the 2017 April 6 observations of \m87.
\autoref{fig:visibility-self-fit} shows a corner plot of the mass-to-distance ratio, spin, spin axis inclination, jet opening angle, spin axis position angle and characteristic radius of the emissivity function ($\theta_g$, $a$, $\theta_{\rm o}$, $\theta_s$, $p.a.$, and $R$) of samples from the posterior.

We cluster the samples on $\theta_g$ into two clusters using a k-means clustering algorithm.
Since $\theta_g$ acts as a scaling parameter on the image, the panels of \autoref{fig:visibility-self-fit} depicting the joint posterior with $\theta_g$ (left most column) show the influence of various modeling parameters on the image size.
Other than $\theta_g$, the parameter that appears to have the most direct effect on the image size is the characteristic radius $R$, as is evident from the characteristic inverse relationship between the two parameters seen in the plot.
A similar trend is seen in Figure 7 of \citet{Palumbo_2022}.

\subsection{Visibility Domain Fit: GRMHD}
\label{sec:visibility-domain-grmhd-fit}

\begin{figure}
\centering
    \includegraphics[width=1.0\linewidth]{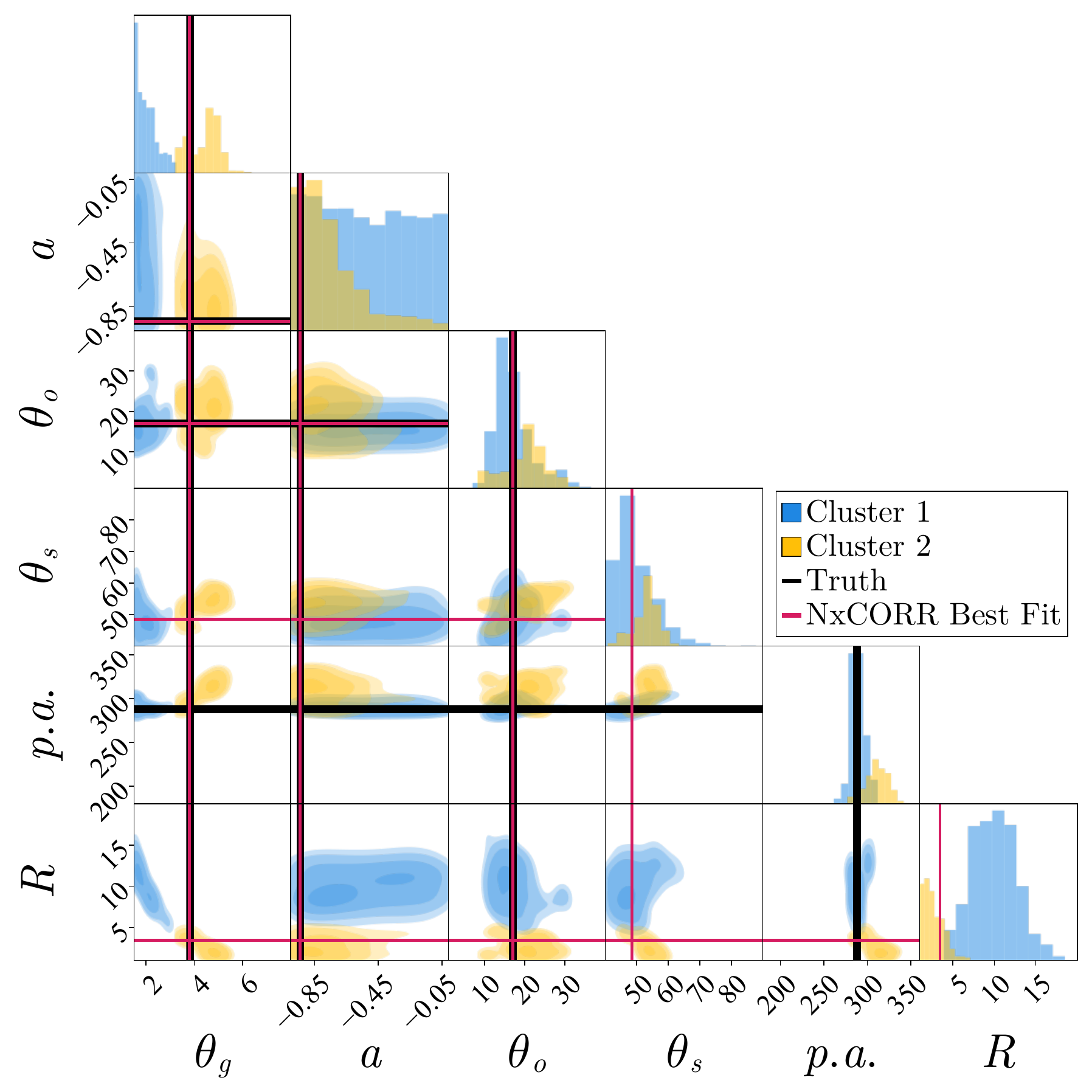}
    \caption{
        Partial corner plot showing the mass-to-distance ratio ($\theta_g$), the black hole spin ($a$), the observer inclination ($\theta_{\rm o}$), the jet opening angle ($\theta_s$), the projected position angle of the spin axis on the observer's screen ($p.a.$), and the characteristic radius of the emissivity function ($R$).
        The true values of the GRMHD simulation are shown with the heavy black lines; the values inferred from by the model in our image domain NxCORR fit of \autoref{sec:image_domain} are shown in red. 
        \autoref{fig:visibility-grhmd-fit-full} shows the full corner plot for this fit. The posterior features two distinct clusters, depicted in blue and yellow, which we isolate using the k-means clustering algorithm.
    }
    \label{fig:visibility-grmhd-fit}
\end{figure}
\begin{figure}
    \centering
    \includegraphics[width=1.0\linewidth]{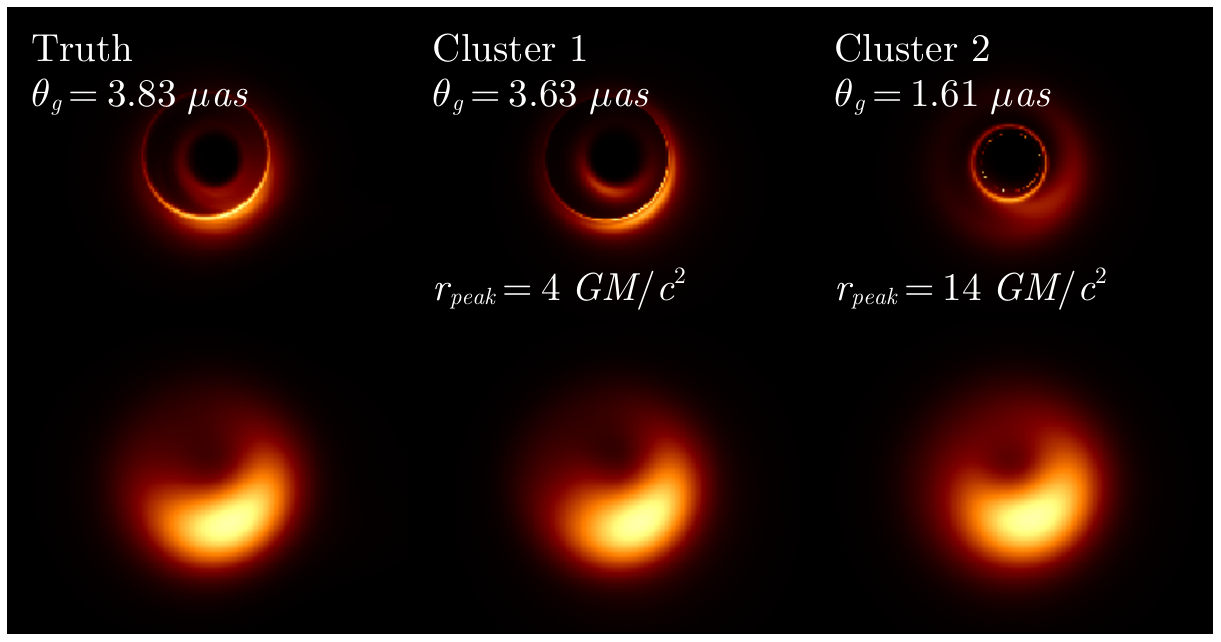}
    \caption{
        True image of the GRMHD simulation (left) with samples from the GRMHD posterior from each mode of the posterior (center and right). The top row shows images at the native resolution used to generate the data; the bottom row shows the same images after being blurred to the nominal EHT resolution of $20\,\mu\text{as}$.
        All images feature a disjoint double ring structure that is not discernible when seen at the EHT nominal resolution.
        }
        \label{fig:grmhd-fit-modes}
\end{figure}

\begin{table}[ht]
\centering
\begin{tabular}{c|cc|cc}
\hline
\hline
\textbf{Params.} & \multicolumn{2}{c|}{\textbf{Cluster 1}} & \multicolumn{2}{c}{\textbf{Cluster 2}} \\
\cline{2-5}
 & \textbf{Low} & \textbf{High} & \textbf{Low} & \textbf{High} \\
\hline
$m_d$ & 1.50 & 2.93 & 3.23 & 5.21 \\
$a$ & -1.00 & -0.06 & -1.00 & -0.28 \\
$\theta_{\rm o}$ & 10$^\circ$ & 30$^\circ$ & 9$^\circ$ & 29$^\circ$ \\
$\theta_s$ & 40$^\circ$ & 62$^\circ$ & 45$^\circ$ & 60$^\circ$ \\
p.a. & 275$^\circ$ & 306$^\circ$ & 283$^\circ$ & 332$^\circ$ \\
$R$ & 4.76 & 15.10 & 1.00 & 4.91 \\
$p_1$ & 0.87 & 9.71 & 0.10 & 9.46 \\
$p_2$ & 4.77 & 7.34 & 3.75 & 6.36 \\
$\chi$ & -140$^\circ$ & -56$^\circ$ & -104$^\circ$ & 32$^\circ$ \\
$\iota$ & 10$^\circ$ & 90$^\circ$ & 37$^\circ$ & 90$^\circ$ \\
$\beta_v$ & 0.36 & 0.90 & 0.18 & 0.82 \\
$\sigma$ & 0.48 & 4.96 & 0.46 & 4.80 \\
$\eta$ & -142$^\circ$ & 93$^\circ$ & -131$^\circ$ & 80$^\circ$ \\
\hline
\end{tabular}
\caption{Highest probability density interval (HPDI), the smallest interval containing $95\%$ of the posterior weights, of modes 1 \& 2 from the posterior samples of our dual cone model fitted to a time-averaged, prograde, SANE GRMHD simulation with \rhigh$=160$.
}
\label{tab:grmhd_hdpi}
\end{table}
In this sub-section, we study the fit of our dual cone model to synthetic observations generated from the prograde, SANE GRMHD model of \autoref{fig:emissivity}.
The main results of this section are summarized in \autoref{fig:visibility-grmhd-fit}, which show the posterior samples for some parameters of interest which we have classified into two distinct clusters on $\theta_g$ with the k-means clustering algorithm. 
The true `on-sky' GRMHD image is shown in the first column of \autoref{fig:grmhd-fit-modes}, while the second and third columns show representative images from each cluster at their true resolution (upper row) and at the EHT nominal resolution of $20\,\mu\text{as}$ (lower row). 
A corner plot of all parameters with their posterior samples can be found in \autoref{fig:visibility-grhmd-fit-full}.

The two clusters differ in which emission features are targetted by either the $n=0$ or $n=1$ photon rings. 
For the first cluster, the decomposition of $n=0$ and $n=1$ emission matches the ground truth image better and is visually similar to the results from \autoref{sec:image_domain}. 
For the second cluster, the $n=1$ emission of the dual cone model focuses on the emission from the jet funnel in the SANE simulation. 
This multi-clustered posterior is directly a result of the finite coverage of the EHT array, as is demonstrated by the lower row of \autoref{fig:visibility-grhmd-fit-full}, where after blurring both clusters appear identical and consistent with the blurred true image.
Most images generated from the posterior samples feature two disjoint rings that are visually consistent with the true image. These two clusters illustrate the expected degeneracy that should occur between the emission radius and black hole mass, \citep[see, e.g.,][]{Palumbo_2022}.

Finally, despite the differences in appearance, both clusters predict a prograde inflow accretion with a significant scale height on the emission surface.
These features are consistent with the image domain analysis of \autoref{fig:emissivity} and with the true simulation.

\begin{figure}[t]
    \centering
    \includegraphics[width=0.475\textwidth]{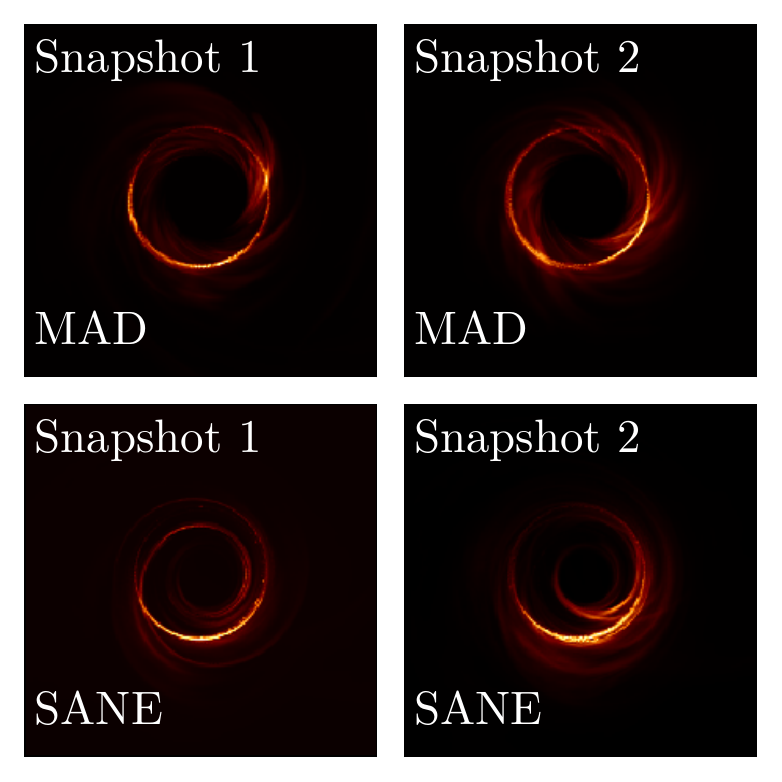}
    \caption{
        Snapshots of an \rhigh$=20$ simulation of a MAD accretion flow (top) and an \rhigh$=160$ simulation of a SANE accretion flow (bottom). These snapshots were used to produce data for the model fits shown in \autoref{fig:visibility-grmhd-snapshot-fit}.
    }
    \label{fig:snapshot-images}
\end{figure} 

As in GRMHD simulations, the accretion flows in LLAGN are composed of turbulent plasma and are observed to be highly variable \citep[see the conclusions of][]{m87V}. This variability means that fits to ``snapshots'' will differ from fits to time-averaged images, with variability introducing a source of misspecification for our model. Because the variability timescale for \m87 is ${\sim}$days, snapshot fits to GRMHD are an appropriate proxy for fits to EHT data from \m87 over a single night.

\begin{figure}[htpb]
    \centering
    \includegraphics[width=0.475\textwidth]{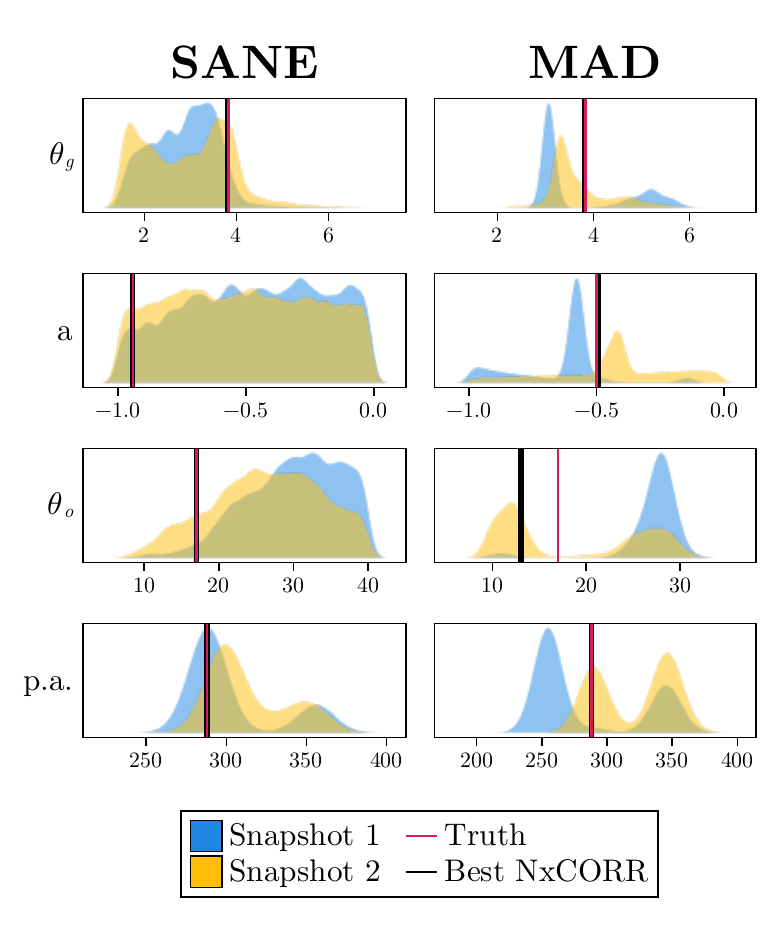}
\caption{
    Marginalized posteriors for snapshot fits to a \rhigh$=160$ SANE simulation (left) and \rhigh$=20$ MAD simulation (right).
    The distributions of two snapshot fits to each simulation are shown in blue and yellow. 
    The parameters shown are the mass to distance ratio ($\theta_g$), spin ($a$), spin axis inclination with respect to the observer ($\theta_o$) and the spin axis position angle (p.a.). 
    For each parameter, the true values and the best NxCORR values that were taken from fits to the time-averaged model in \autoref{sec:image_domain} are shown with red and black vertical lines, respectively.
    }
    \label{fig:visibility-grmhd-snapshot-fit}
\end{figure}

To assess the effects of this time-variable structure, we fitted snapshots of both MAD and SANE simulations using our dual-cone model (see \autoref{fig:snapshot-images}). \autoref{fig:visibility-grmhd-snapshot-fit} shows density plots for these fits.  These fits suggest that the systematic errors expected in our model could be dependent on the underlying accretion flow, with fits to snapshots of the SANE model showing less variance than the MAD fits. 
Notably, both snapshot fits to the SANE simulation contain the values of the best NxCORR representative model in the bulk of their posteriors. 
The MAD fits, in contrast, have a larger variance between the two posteriors, with only one posterior containing all values of the best NxCORR representative model in its bulk.



\subsection{Visibility Domain Fit: EHT Data}
\label{sec:visibility-domain-data-fit}

\begin{figure}[htpb]
    \includegraphics[width=0.47\textwidth]
    {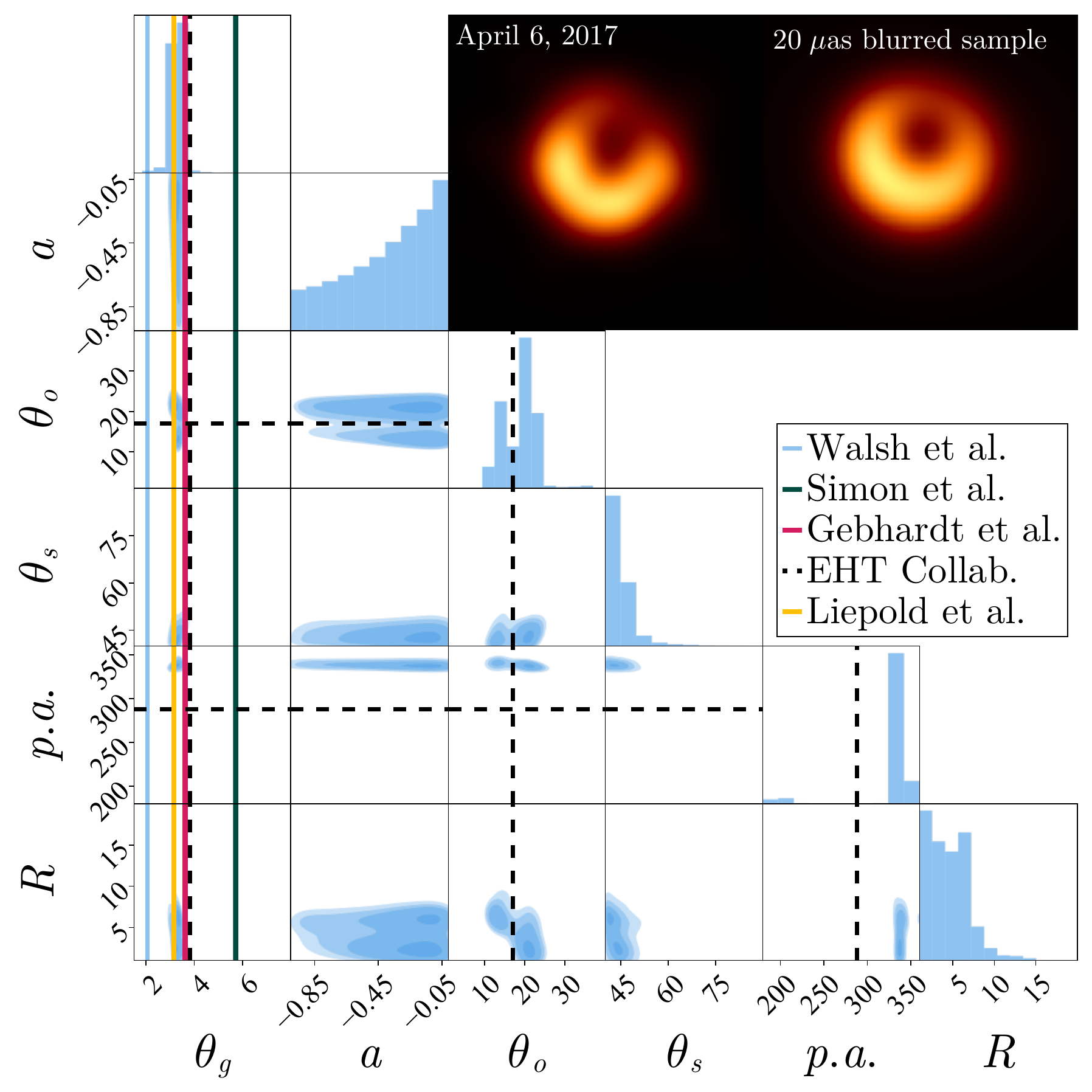}
    \caption{
        Partial corner plot for a dual cone fit to the April 6th 2017 EHT observations of \m87. 
        The shown parameters are the mass-to-distance ratio ($\theta_g$), the black hole spin ($a$), the observer inclination ($\theta_{\rm o}$), the jet opening angle ($\theta_s$), the projected position angle of the spin axis on the observer's screen ($p.a.$), and the characteristic radius of the emissivity function ($R$). 
        Vertical lines show other mass measurements of \m87 in red \citep{Gebhardt}, blue \citep{Walsh}, black dotted \citepalias{EHTC_M87_VI}, yellow \citep{Liepold_2023}, and green \citep{Simon}. 
        We also show the measured position angle and inclination of the large-scale jet in \m87 \citep{Walker}. 
        The inset images show the consensus April 6th image of \m87 reported by the EHTC and a random sample drawn from the posterior, blurred with a 20\,$\mu{\rm as}$ FWHM Gaussian beam. 
        Both images have their peak brightness in the southeastern quadrant.
    }
    \label{fig:visibility-data-fit}
\end{figure}
\begin{table}[ht]
\centering
\begin{tabular}{c|cc|cc|cc}
\hline
\hline
\textbf{Params.} & \multicolumn{3}{c}{\textbf{Low}} & \multicolumn{3}{c}{\textbf{High}} \\
\hline
$\theta_g$ & \multicolumn{3}{c}{2.84} & \multicolumn{3}{c}{3.75} \\
$a$ & \multicolumn{3}{c}{-0.90} & \multicolumn{3}{c}{-0.01} \\
$\theta_{\rm o}$ & \multicolumn{3}{c}{11$^\circ$} & \multicolumn{3}{c}{24$^\circ$} \\
$\theta_s$ & \multicolumn{3}{c}{$40^\circ$} & \multicolumn{3}{c}{$56^\circ$} \\
p.a. & \multicolumn{3}{c}{200$^\circ$} & \multicolumn{3}{c}{347$^\circ$} \\
$R$ & \multicolumn{3}{c}{1.00} & \multicolumn{3}{c}{8.46} \\
$p_1$ & \multicolumn{3}{c}{0.71} & \multicolumn{3}{c}{9.99} \\
$p_2$ & \multicolumn{3}{c}{1.47} & \multicolumn{3}{c}{7.27} \\
$\chi$ & \multicolumn{3}{c}{35$^\circ$} & \multicolumn{3}{c}{140$^\circ$} \\
$\beta_v$ & \multicolumn{3}{c}{0.08} & \multicolumn{3}{c}{0.55} \\
$\iota$ & \multicolumn{3}{c}{10$^\circ$} & \multicolumn{3}{c}{49$^\circ$} \\
\cline{2-7}
 & \multicolumn{2}{c}{\textbf{Cluster 1}} & \multicolumn{2}{c}{\textbf{Cluster 2}} & \multicolumn{2}{c}{\textbf{Cluster 3}}\\
 \cline{2-7}
 & \textbf{Low} & \textbf{High} & \textbf{Low} & \textbf{High} & \textbf{Low} & \textbf{High}\\
$\sigma$ & 1.73 & 5.00 & 1.29 & 3.95 & 2.98 & 5.00\\
$\eta$ & -19$^\circ$ & 27$^\circ$ & 151$^\circ$ & 180$^\circ$ &-180$^\circ$ & -149$^\circ$\\
\hline
\end{tabular}
\caption{95\% Highest probability density interval (HPDI)  of our dual cone model fitted to closure quantities of \m87 taken from the EHTC observations on April 6th, 2017.
The posterior features 3 modes in $\eta$.
These modes have largely similar HPDI except for their spectral indices, which are shown separately.
Though the HPDI reports a broad interval for the inferred position angle of the projected spin axis, the distribution has two separate clusters in $p.a.$, as can be seen in \autoref{fig:visibility-self-fit}, which excludes the truth.
}
\label{tab:HDPI_EHT}
\end{table}

We also used our model to fit EHT observations on \m87. In this case, the limitations of model errors are unknown, but we can still evaluate the ability of the the model to reproduce actual measurements. Nevertheless, while we have shown that our model accurately represents time-averaged GRMHD with reasonable parameters, realistic data will include degrees of freedom not included in the dual cone model. For instance, GRMHD simulations show significant structural turbulence in snapshots. These turbulent structures are not captured in our dual cone model and may introduce systematic errors that are not identified in the tests we have performed.

We conduct our analysis on the April 6th observations of \m87, following \citetalias{EHTC_M87_V}. 
\autoref{fig:visibility-data-fit} summarizes our results; the full posterior samples can be found in \autoref{fig:visibility-data-fit-full}. The $95\%$ HPDI for the parameters of our model in \autoref{tab:HDPI_EHT}.
We infer a relatively narrow posterior for the mass-to-distance ratio $\theta_g\in(2.84,3.75)$ from our fits, consistent with the stellar mass estimates of \citet{Gebhardt,EHTC_M87_VI,Liepold_2023}.
In contrast to $\theta_g$, our spin inference is weakly constrained, a result that is consistent with the results of the previous self-fit and GRMHD fit.
Our inability to constrain the black hole spin with our model, even when performing self-fits, suggests that spin measurements with the 2017 EHT array are infeasible.
Furthermore, restricting the the priors on the emission geometry would not improve the spin constraints significantly since the emission parameters are uncorrelated with spin for Stokes I.
The lack of correlation of spin, at EHT resolution, with any of the model parameters is evident from the structure of the posteriors in the spin columns of \autoref{fig:visibility-grhmd-fit-full} and \ref{fig:visibility-data-fit-full}, and holds even in the self-fit of 
\autoref{fig:visibility-self-fit-full}.

A spin measurement that relies on Stokes I quantities will likely require improved resolution or sensitivity to dynamical signatures to discern the effects of spacetime from accretion. 
Future proposed VLBI arrays, such as extensions of the EHT to space with the Black Hole Explorer, will aim to measure the $n=1$ photon ring directly, providing an avenue to Stokes I spin measurements. At the resolution of the EHT, including polarization data may improve the prospects of spin constraints due to frame-dragging effects on the polarization pattern \citep{Palumbo_2020,Palumbo_2022b,Chael_2023} and improved astrophysical constraints \citepalias[e.g.,][]{EHTC_M87_VIII,EHTC_M87_IX,EHTC_SgrA_VIII}.

Although we are unable to constrain the magnitude of the spin from our model, we measure a spin axis inclination that is consistent with the direction of the large-scale jet, $\theta_{\rm o}\in(11^\circ,24^\circ)$ \citep{Walker}. In contrast, the EHTC analysis of \m87 was performed by assuming that the spin axis is aligned with the large scale jet. 
Misaligned disks are, however, believed to be a generic feature of many AGN, with some authors arguing for a tilted flow in the \m87 system from the wobbling of the large scale jet \citep[e.g.,][]{m87tilt}, so this independent measurement of inclination is both encouraging for the EHTC conclusions and non-trivial.  

We emphasize, however, that our model makes this inference under the assumption that the spin axis is aligned with the symmetry axis of the horizon-scale accretion flow.
One can expect an alignment of the horizon scale accretion flow with the spin axis of the black hole from Bardeen-Peterson effect which applies a torque to misaligned viscous flows \citep{BardeenPeterson, Membrane}.
GRMHD simulations often do not include a viscosity term in their evolution, but the presence of the magnetic field appears to introduce an effective viscosity in the dynamics that has been seen to align flows in simulations that feature tilted disks.
\citep[see for example the tilted disk simulations of][]{LiskaTiltedDisk} .
This effect is is typically seen to occur within a few tens of gravitational radii of the hole, where the bulk of the emission observed by the EHT originates.

Despite the forced alignment, the presence of a tilted flow can still leave imprints on images of GRMHD through its effect on the jet, though some of these features are suppressed at the $230$\,GHz frequencies observed by the EHT \citep{KoushikTiltedDisk}.
In particular, the shape of the photon ring appears largely unaffected.
It is thus reasonable to expect that our model should still be sensitive to the inclination of the spin of the black hole, rather than the inclination of the large scale accretion.

We also infer a projected position angle of the spin axis that points more northerly than would be expected if it was aligned with the large-scale jet (see \autoref{fig:visibility-data-fit}). Studies on additional observing epochs will be necessary to clarify whether this could be driven by transient emission structure. 

The design of our model allows us to infer the 3D emission geometry of our source.
We infer an emission profile with characteristic radius $R\in(1,8.46)$, inner exponent $p_1\in(0.71,9.99)$, and outer exponent $p_2\in(1.47,7.27)$.
From \autoref{eqn:double_power_law}, we can infer a peak radius,
\begin{align}
    R_{\text{peak}}
        &=R\left(\frac{p_1}{p_2}\right)^{\frac{1}{p_1+p_2}},
\end{align}
which corresponds to a peak radius in the range $R_{\text{peak}}\in(0.84,10.40)$ that is similar to GRMHD simulations \citep{Dexter,m87V,Emami_2023}. Our fitted emissivity parameters suggest that the emission geometry has a half opening angle of $\theta_s\in(40^\circ,56^\circ)$. The inferred accretion flow is inferred to be slow-to-moderately relativistic and retrograde, $\beta_v\in(0.08,0.55)$, although we are unable to constrain whether it is inflowing or outflowing. Finally, we are unable to constrain aspects of the spectral index and magnetic field geometry tightly. 
There are many opportunities to verify these broad conclusions through additional data, including EHT observations of \m87 in other years \citepalias{EHTC_M87_2018}, as well as fits using multi-frequency data and polarization information \citepalias{EHTC_M87_VII,EHTC_M87_VIII,EHTC_M87_IX}.

\section{Summary}
\label{sec:discussion}
We have introduced a semi-analytic model for emission concentrated within a conical region near a black hole. 
Despite its simplicity, we have demonstrated that this model provides excellent approximations to a wide variety of time-averaged GRMHD simulations. 
Moreover, our model successfully achieves the goal of \emph{photogrammetry} by capturing the properties of both the black hole and the surrounding emission when fit to high-resolution images or interferometric data. 

Our model is highly efficient, both in terms of the expense to generate images (e.g., we are able to generate $400\times400$ pixel images in ${\approx}10\,\mu\text{s}$ with the integrated GPU on a laptop with the Apple M1 architecture) and in providing a low-dimension representation of horizon-scale images from LLAGN. 
For instance, our dual cone model has only 13 parameter; for comparison, the \texttt{xs-ringauss} geometric model used by the the EHT for their mass measurement of \m87 requires 26 parameters \citepalias{EHTC_M87_VI}. 
A further benefit of dual cone model fits is that they directly measure physically relevant parameters (e.g., the black hole mass and spin, and properties of the emitting plasma), while the geometric model fits require an additional layer of analysis to interpret their parameters (e.g., the $\alpha$-calibration procedure used in the EHTC; see \autoref{sec:introduction}). 

We show that our model is capable of reproducing time-averaged images from a wide class of GRMHD simulations. 
Unlike previous semi-analytic models that were restricted to equatorial emission, our model reproduces characteristic image morphologies that are seen in both SANE and MAD GRMHD simulations, they produce emission geometries that are broadly consistent with the true emission geometry of the underlying simulation, and are often able to discern the fluid direction (prograde or retrograde) from images.



We assess the suitability of our model for interpreting VLBI measurements using a series of tests that fit mock EHT observations of simulated images. We explore ``self-fits'' of the dual cone model (\autoref{sec:visibility-domain-self-fit}) as well as fits to mock observations of both time-averaged and snapshot GRMHD images (\autoref{sec:visibility-domain-grmhd-fit}). In both cases, the model fits give accurate mass measurements but do not meaningfully constrain the black hole spin, although we do see more significant biases from model misspecification in snapshot fits. These tests reinforce the EHTC conclusions that the spin cannot be measured from the current data without strong model assumptions. 


We also used our dual cone model to fit real EHT observations of \m87. As for the snapshot fits, we expect additional systematic errors from time-variable image structure (in addition to systematic errors from imperfections in the model specification). Nevertheless, these fits give estimates for the mass and inclination that are consistent with EHT measurements. However, we do see some evidence for systematic errors in the fits, including a position angle for the black hole spin axis that differs from the direction of the large-scale jet. 

Future extensions of the model could include linear and circular polarization, both of which provide important constraints in EHT analyses. They could also relax the assumption of axisymmetry, to directly model time-variable structures. Alternatively, comparison of model fits over many gravitational timescales in \m87 (e.g., over multiple years of EHT observations) will provide an empirical estimate for the systematic uncertainties associated with model misspecification and may provide guidance into what physical parameters can be confidently inferred from these fits.

\begin{acknowledgments}
\section{Acknowledgements}
We thank Koushik Chaterjee, Razieh Emami, Charles Gammie, Sara Issaoun, Prashant Kocherlakota, Ramesh Narayan, Dominic Pesce, Angelo Ricarte and George Wong for their insightful discussions.
We acknowledge financial support from the Brinson Foundation, the Gordon and Betty Moore Foundation (GBMF-10423), and the National Science Foundation (AST-2307887, AST-1935980, and AST-2034306). 
This work was supported by the Black Hole Initiative, which is funded by grants from the John Templeton Foundation (Grant \#62286) and the Gordon and Betty Moore Foundation (Grant GBMF-8273) - although the opinions expressed in this work are those of the author(s) and do not necessarily reflect the views of these Foundations.
This work used the RCAC Anvil Cluster at Purdue University through allocation PHY230186 from the Advanced Cyberinfrastructure Coordination Ecosystem: Services \& Support (ACCESS) program \cite{Access}, which is supported by National Science Foundation grants \#2138259, \#2138286, \#2138307, \#2137603, and \#2138296.
\end{acknowledgments}
\appendix
\section{Model Definition}
\subsection{Interpreting Model Parameters}
\begin{figure*}[t]
    \centering
    \includegraphics[width=0.9\textwidth]{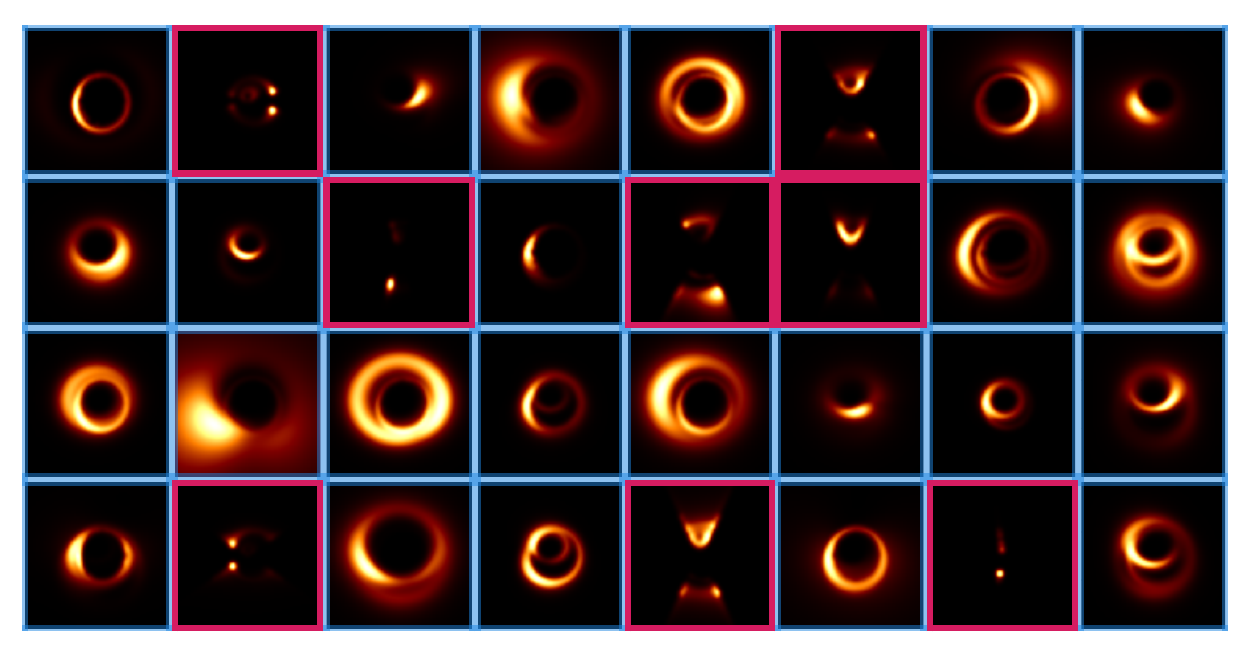}
    \caption{
    Random sample of images from the dual-cone model seen by a distant observer.
    All images shown are comprised of $n=0$ and $n=1$ sub images.
    In all images, the mass-to-distance ratio is the same, the observer inclination is $30^\circ$, the spin is $-1$, and the projected spin axis is oriented vertically on the observer's screen.
    We have classified the images into two categories, `inside the cone' (blue) and `outside the cone' (red).
    These are defined respectively by images generated with $|\cos\theta_s| < |\cos\theta_o|$ in the case of `inside the cone' images or $|\cos\theta_s| > |\cos\theta_o|$ in the case of `outside the cone' images.
    }
    \label{fig:random}
\end{figure*}

Much of the interpretation of the dual-cone model parameters are gleaned from studies of the model structure. 
Here, we summarize some key features of the model to provide additional background and intuition.
\autoref{fig:random} shows an example of what images of the model can look like for a random sample of emission parameters.
The mass, spin, observer inclination, and position angle are fixed.
Two classes of images are generically formed whose classification depends on the observer's relative orientation with respect to the emission cone.
These two classes fall broadly into the categories of `inside the cone' or `outside the cone' and matches that of \citet{Papoutsis}.
Observers that views the black hole with a configuration that satisfies $|\cos\theta_o|>|\cos\theta_s|$ sees images in the `inside the cone' category, while observers viewing the black hole with a configuration $|\cos\theta_o|<|\cos\theta_s|$ sees images in the `outside the cone category. 
\autoref{fig:random} shows example images from each class, all featuring both $n=0$ and $n=1$ emission.
Some of the `inside the cone' images depict only a single ring-like feature instead of multiple, which is emblematic of the flexibility of our anisotropic synchrotron emission model allowing for the peak intensity of the $n=0$ image to be greater than that of $n=1$, despite the fact that the dual-cone model is optically thin.
Although our parameterization is axisymmetric, our emissivity model generically forms asymmetric emission structures in the bulk spacetime.

\begin{figure*}[t]
    \centering
    \includegraphics[width=\textwidth]{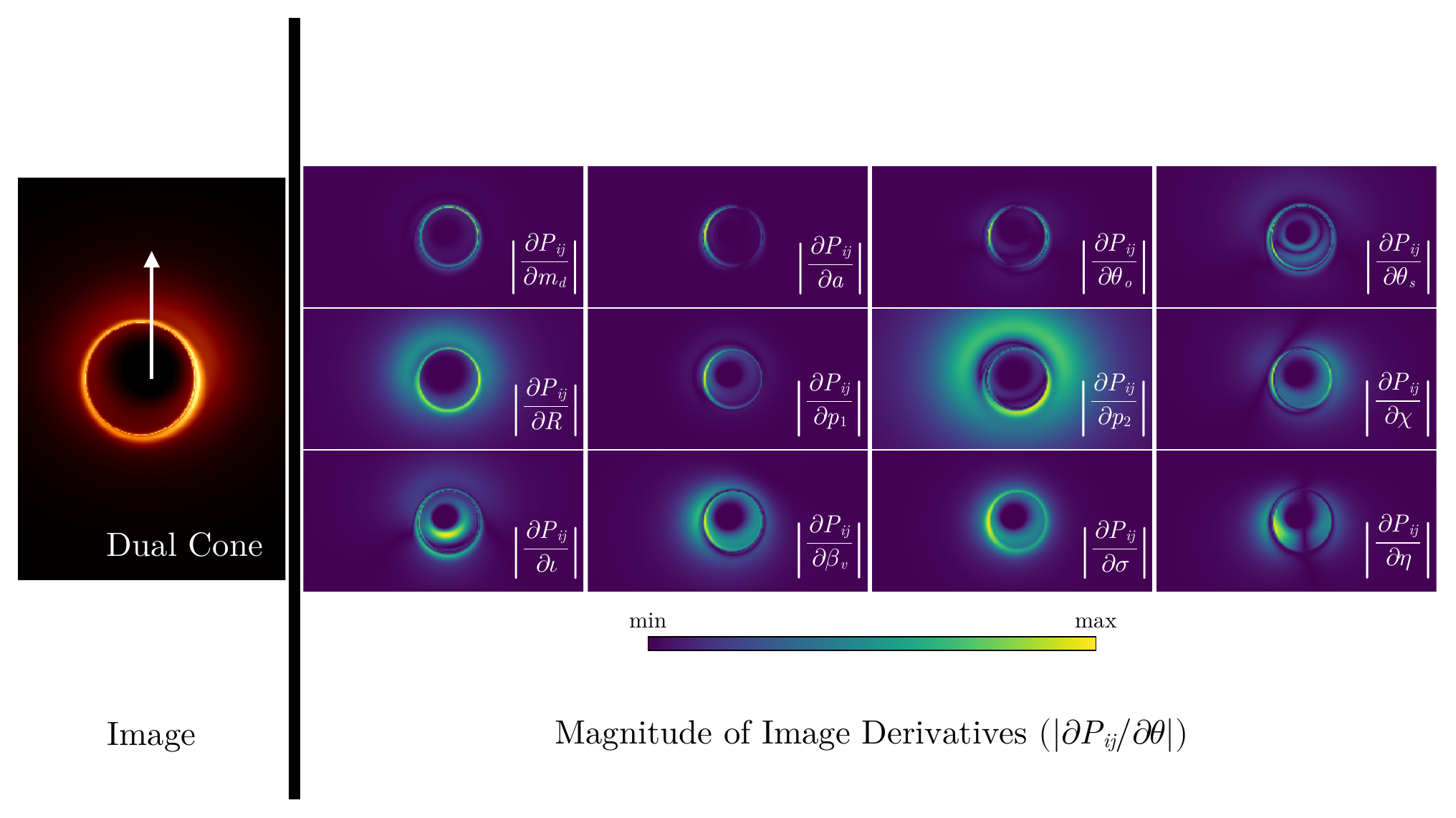}
    \caption{
        Assessing the influence of model parameters on various image features.
        The leftmost panel shows a image of one of the representative dual-cone models of \autoref{sec:image_domain}.
        The direction of the spin is projected onto the image with a white arrow.
        The grid of panels to the right shows the magnitude of the partial derivatives of each image pixel with respect the parameters of the dual-cone model.
        The color range in each of the derivative panels is re-scaled to its minimum and maximum pixel value. 
    }
    \label{fig:gradients}
\end{figure*}

We give some intuition on how various parameters can affect features in the images of the model in \autoref{fig:gradients}.
This plot depicts gradients of the image pixels with respect to the model parameters when evaluated at one of the best fit images in \autoref{sec:image_domain}.
The left most panel shows the true image with the direction of the projected spin axis super imposed as white arrow.
The parameters can broadly be categorized as, spacetime dependent parameters -- the mass-to-distance ratio ($\theta_g$), spin ($a$), and inclination ($\theta_o$) -- which are seen to have large gradients in only in the photon ring region, and emissivity and accretion parameters that largely affect features in either only the $n=0$ subimage, or both $n=0$ and $n=1$ sub images.

The image structure of spacetime dependent gradients is consistent with the expected action of these parameters on the appearance of the photon ring.
\citet{Papoutsis} have suggested, for example, that a shift in the photon ring with respect to the projected the spin axis is likely an observable sensitive to spin, since the the $n=0$ sub-image is largely insensitive to spin.
This effect can be seen in our spin derivative panel (first row, second column of the image gradients), which features a large gradient for emission in the $n=1$ sub image and a relatively suppressed gradient for $n=0$. 
The magnitude of the gradient is higher in regions of the $n=1$ sub image that are away from the spin axis.
The asymmetry in the sensitivity of the $n=1$ feature to spin is indicative of the effect of spin on the photon ring's displacement.
An other example is of the gradient with $m_d$ (first row first column) which is also strongest in the photon ring region.
The $m_d$ parameter acts as a radial scaling of the blackhole, a fact that is exemplified by the relatively uniform gradient around the $n=1$ ring in the $m_d$ panel.

Some of the emissivity parameters of interest are those which control the location of the peak emission $R$, $p_1$ and $p_2$. 
The magnitude of the derivative with respect to $p_1$ are peaked within a compact region, interior to the peak emission, while those of $p_2$ are peaked exterior to that region. 
These results are consistent with the definition of our emissivity envelope in \autoref{eqn:double_power_law}.
Some of the accretion parameters, including fluid speed ($\beta_v$) and magnetic field azimuthal orientation ($\eta$) largely do not influence the $n=1$ emission structure. 
We note that \autoref{fig:gradients} demonstrates that the ``inner shadow'' \citep{Chael_2021} is sensitive to the jet opening angle ($\theta_s$) where the jet base of the forward and rear jets are seen highlighted in the gradient with respect to $\theta_s$. 

\subsection{Transformations of Quantities to Boyer-Lindquist Coordinates}

\autoref{eqn:tetrads} can be used to map vectors in the ZAMO frame to vectors in the Boyer-Lindquist frame.
Performing this transformation on the fluid velocity gives,
\begin{align}
   u^\mu 
    &=u^{(m)}e_{(m)}{}^{\mu}=
   \beta_v\begin{pmatrix}
        0,
        \sqrt{\frac{\Delta}{\Sigma_s }}  \cos \chi ,
        0,
        \sqrt{\frac{\Sigma_s}{\Xi_s}} \frac{\sin\chi}{\sin\theta} ,
   \end{pmatrix}
\end{align}
which shows that the intuition that $\chi$ controls the direction of the fluid flow carries over from the fluid frame to the global Boyer-Lindquist frame.

We can also transform the magnetic field vector in a similar fashion.
Since there is no electric field in the fluid frame, a magnetic field 4-vector can be constructed from $B^{m'}$ to find the magnetic field in the global Boyer-Lindquist frame \citep{harm_2003},
\begin{align}
    b^{m'}
        &=\frac{1}{2}\epsilon^{m'n'o'p'}u_{n'}F_{o'p'}.
\end{align}
From this definition, it follows that $B^{m'}=b^{m'}$ in the fluid frame.
The equivalent 4-vector in the Boyer-Lindquist Frame is then,
\begin{align}
    b^{\mu}
        &=e^{\mu}{}_{(m)}\Lambda^{(m)}{}_{(n')}b^{(n')},
\end{align}
where $\Lambda^{(m)}{}_{(n')}=\left(\Lambda^{(n')}{}_{(m)}\right)^{-1}$ is the inverse Lorentz transformation.
The components of $b^\mu$ are,
\begin{subequations}
\begin{align}
    b^t
        &=\beta  \gamma  \sqrt{\frac{\Xi }{\Delta  \Sigma }} (\sin (\theta ) \sin (\iota )
   \cos (\eta -\phi )+\cos (\theta ) \cos (\iota ))\\
b^r
    &= \sqrt{\frac{\Delta }{\Sigma }} \Big(\frac{1}{8} \sin (\iota ) \left(2 (\gamma +3)
   \cos (\eta )-(\gamma -1) \left(-4 \sin ^2(\theta ) \cos (\eta -2 \phi )+\cos
   (\eta -2 \theta )+\cos (\eta +2 \theta )\right)\right) \\
   &\hspace{10cm} + (\gamma -1) \sin (\theta
   ) \cos (\theta ) \cos (\iota ) \cos (\phi )\Big)\nonumber\\
b^\theta
    &=-\frac{(\gamma -1) \sin (2 \theta ) \sin (\iota ) \cos (\eta -\phi )+\cos (\iota )
   ((\gamma -1) \cos (2 \theta )+\gamma +1)}{2 \sqrt{\Sigma }}\\
b^\phi
    &=\beta  \gamma  \omega  \sqrt{\frac{\Xi }{\Delta  \Sigma }} (\sin (\theta ) \sin
   (\iota ) \cos (\eta -\phi )+\cos (\theta ) \cos (\iota ))\\
   &+\frac{1}{8} \csc
   (\theta ) \sqrt{\frac{\Sigma }{\Xi }} \Big(\sin (\eta ) \sin (\iota ) (\gamma  \cos
   (2 (\theta +\phi ))+(\gamma -1) \cos (2 (\theta -\phi ))-2 (\gamma -1) \cos (2
   \theta )-2 \gamma  \cos (2 \phi )+2 \gamma \nonumber\\
   &-\cos (2 (\theta +\phi ))+2 \cos (2
   \phi )+6)+8 (\gamma -1) \sin (\theta ) \sin (\phi ) (\cos (\eta ) \sin (\theta )
   \sin (\iota ) \cos (\phi )+\cos (\theta ) \cos (\iota )) \Big). \nonumber
\end{align}
\end{subequations}

\subsection{Image Indexing}
\label{sec:image_index}
\begin{figure*}[hptb]
    \centering
    \includegraphics[width=\textwidth]{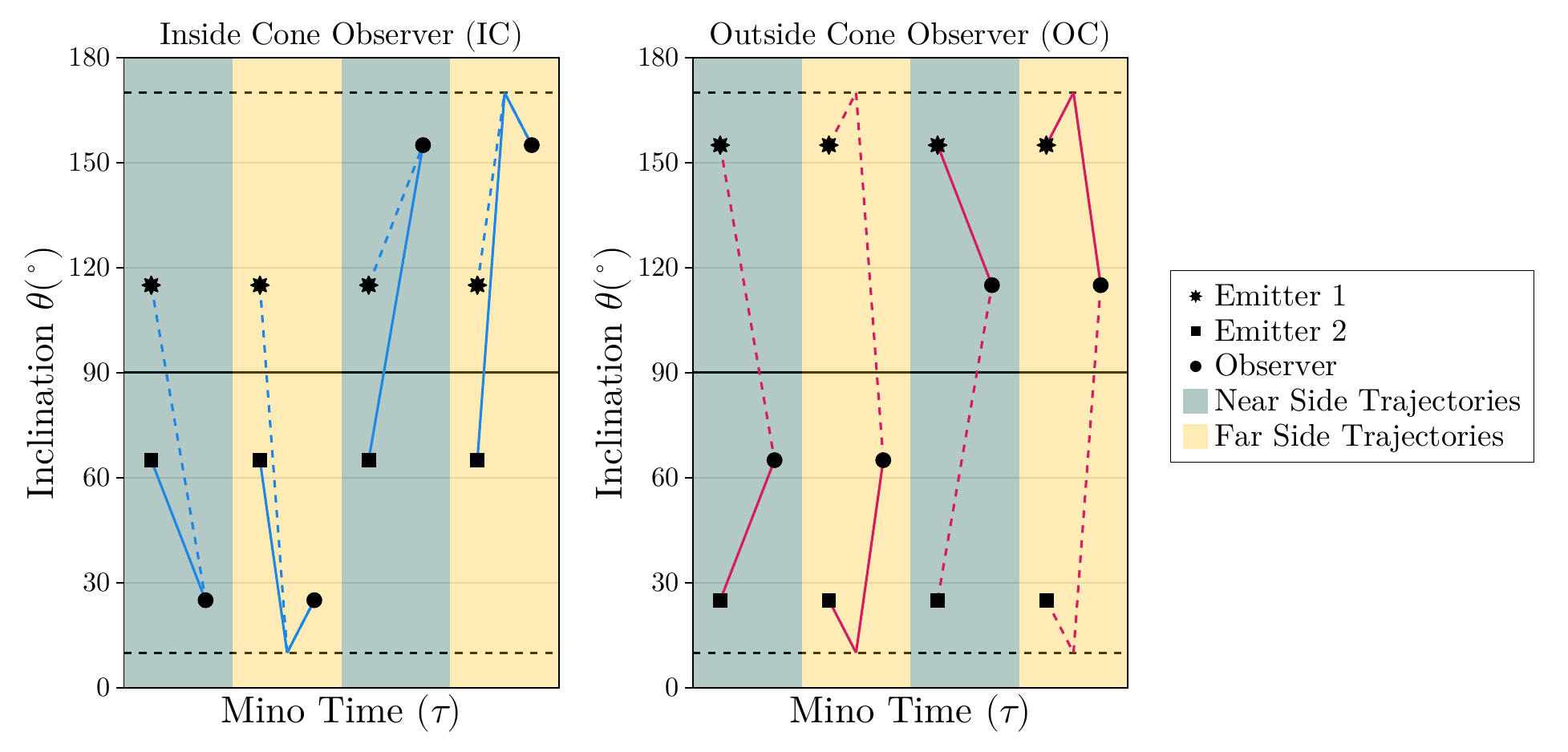}
    \caption{Schematic diagram showing the types of trajectories that can make up the $n=0$ sub-image. 
    Multiple configurations are shown.
    The left panel shows the trajectories for observer and emitter configurations where the observer is inside the cone (IC), while the right panel shows configuration for where the observer is outside the cone (OC).
    The green highlighted trajectories originate from emitters that are on the side of the cone near to the observer, while those highlighted in orange originate from the side far from the observer.
    }
    \label{fig:0-indexing}
\end{figure*}
The Kerr space time naturally gives rise to strong lensing effects where an object will produce infinitely many images when seen by a distant observer \citep[see, e.g.,][]{1973ApJ...183..237C,Ohanian}.
A prescription for indexing the infinity of images formed from a single object is then useful when modelling observables. 
Previous works have focused on schemes which index images based on the number of equatorial crossings in a photon trajectory \citep[see, e.g.,][]{GrallaLupsasca,MJohnsonRing},
but such definitions are ambiguous for models with off-equatorial emission.
Here we define a generic strategy for indexing null-geodesic trajectories in the Kerr space time.

An observer at infinity sets up a screen with coordinates,
\begin{align}
    \alpha =-\frac{p^\phi}{p^r}
        &= -\frac{\lambda}{\sin\theta_{\rm o}}\\
    \beta =  -\frac{p^\theta}{p^r}
        & = -\sqrt{\Theta(\theta_{\rm o}, \lambda, \eta)}\label{eqn:Bardeen_coordinates}
\end{align}
where 
\begin{align}
    \Theta(\theta)
    =\sqrt{\eta+a^2\cos^2\theta-\lambda^2\cot^2\theta},
\end{align}
is the angular potential of a null geodesic.
Every pixel on the screen of an asymptotic observer can be associated with a unique  null ray of the Kerr exterior.
This fact allows for a local indexing scheme which orders pixels by the total affine parameter or the Mino time of the trajectories that generated them.
We note that a indexing scheme based on Mino time is equivalent to one performed on affine time since the two are related through the relationship,
\begin{align}
   d\tau &= \frac{d\lambda}{\Sigma},
\end{align}
where $\Sigma \geq 0$.

We define the Mino-time for an $n^\text{th}$ indexed trajectory which connects a pixel to a bulk location.
We note that the Mino time can be expressed in terms of quadratures of $\theta$ as,
\begin{align}
    \Delta\tau 
        = \fint_{\theta_s}^{\theta_{\rm o}}\frac{\Sigma}{\pm p_\theta(\theta)}d\theta
        &= \fint_{\theta_s}^{\theta_{\rm o}}\pm\frac{d\theta}{\sqrt{\Theta(\theta)}},
        \label{eqn:minotime-theta}
\end{align}
where,
\begin{align}
    \frac{d\theta}{\sqrt{\Theta}}
        &=\frac{d\theta}{\sqrt{\eta+a^2\cos^2\theta-\lambda^2\cot^2\theta}}\\
        &=\frac{1}{2}\frac{d u}{\sqrt{u}\sqrt{\eta-(\eta+\lambda^2-a^2)u+a^2u^2}}\label{eqn:dtau_du},
\end{align}
with $u=\cos^2\theta$, and $\theta_s$ and $\theta_{\rm o}$ being the emission and observation Boyer-Lindquist inclination of the photon.
The symbol $\fint$ of \eqref{eqn:minotime-theta} indicates a path integral.

Any trajectory of \eqref{eqn:minotime-theta} can be classified as `Ordinary' or `Vortical'.
The classification is based off the fact that any point along the photon's trajectory must satisfy the constraint, $\Theta(\theta)\geq 0$, which implies that the rays have turning points at,
\begin{align}
    u_{ \pm}=\triangle_\theta \pm \sqrt{\triangle_\theta^2+\frac{\eta}{a^2}}, \quad \triangle_\theta=\frac{1}{2}\left(1-\frac{\eta+\lambda^2}{a^2}\right).
\end{align}
The classification of trajectories into `Ordinary' or `Vortical' then follows from the nature of the roots  \citep[see][]{GrallaLupsasca}, where geodesics that are said to undergo `Ordinary' motion satisfy,
\begin{align}
    \arccos(\sqrt{u_+})<\theta<\arccos(-\sqrt{u_+}).
\end{align}
while those that undergo `Vortical Motion' satisfy
\begin{align}
    \arccos(-\sqrt{u_-})<\theta<\arccos(-\sqrt{u_+})\quad\text{or}\quad\arccos(\sqrt{u_+})<\theta<\arccos(\sqrt{u_-}),
\end{align}

Our goal of this section is relate the length of the trajectory of a photon to an index, $n$, that  labels the sub-image which the photon generates.
To aid in this analysis, we define the integrals,
\begin{align}
    G_{\theta_s}
        &=\frac{1}{2}\int_{0}^{\cos(\theta_s)}\frac{d u}{\sqrt{u}\sqrt{\eta-(\eta+\lambda^2-a^2)u+a^2u^2}},\\
    G_{\theta_{\rm o}}
        &=\frac{1}{2}\int_{0}^{\cos(\theta_{\rm o})}\frac{d u}{\sqrt{u}\sqrt{\eta-(\eta+\lambda^2-a^2)u+a^2u^2}},\\
    \hat G_\theta  
        &=2\begin{cases}
        \int_{\arccos\left(\sqrt{u_+}\right)}^{\arccos\left(-\sqrt{u_+}\right)}\frac{d u}{\sqrt{u}\sqrt{\eta-(\eta+\lambda^2-a^2)u+a^2u^2}},&\text{Ordinary}\\
        \int_{\arccos\left(\mp\sqrt{ u_-}\right)}^{\arccos\left(-\sqrt{u_+}\right)}\frac{d u}{\sqrt{u}\sqrt{\eta-(\eta+\lambda^2-a^2)u+a^2u^2}},&\text{Vortical}
        \end{cases}\label{eqn:half_orbit}.
\end{align}
These will be useful since the total Mino time can be written as linear combination of these functions.
In general, the total Mino time of the $n^{\text{th}}$ image is,
\begin{align}
    \Delta\tau_n = n\hat G_\theta +\Delta\tau_0,\label{eqn:minotime_n}
\end{align}
which is dependent on the emission and observation inclinations, as well as the pixel on the observer's screen that the photon lands on.

Refining \eqref{eqn:minotime_n} into a definition requires specifying the Mino time of the $0^{th}$ image, $\Delta\tau_0$.
Our approach is to define $\Delta\tau_0$ from the observational properties of a curve $\gamma_\phi=(t, r(\phi), \theta_s, \phi)$, which is taken to be closed in bulk space. 
$\gamma_\phi$ will thus result in an image of a closed curve on the observer's screen that will either be intersected by the $\beta=0$ line, or lie entirely to one side of it.
We use the fact that $\gamma_\phi$ can be thought of the intersection of a rigid cone and a rigid 2D hypersurface, to define the
situation where the image of $\gamma_\phi$ intersects $\beta=0$ as the case where the observer is positioned `inside the cone' (IC), while the the other case will be for an observer positioned `outside the cone' (OC).
We have chosen this classification to be consistent with that of \citet{Papoutsis}, and will quantify it through the definition of a variable, 
\begin{align}
    \text{isincone}
    =
    \begin{cases}
        \text{true},&\lvert\cos\theta_{\rm o}\rvert> \lvert\cos\theta_s\rvert,\;\text{inside the cone (IC)}\\
        \text{false},&\lvert\cos\theta_{\rm o}\rvert< \lvert\cos\theta_s\rvert,\;\text{outside the cone (OC)}
    \end{cases}
\end{align}

One observational distinction that can be made about the emission of $\gamma_\phi$ is whether portions of it originated from sections of the curve that lie on the `far side' of the cone, or on the `near side' with respect to the observer.
This distinction allows us to, define $\Delta\tau_0$ in the following way for IC observers as:
\begin{align}
    \left.\Delta\tau_0\right\rvert_{\text{IC}}=
    \begin{cases}
        G_o+(-1)^{\nu_{\text{indir}}}G_s&, \text{nearside}\\
        \hat G -G_o+(-1)^{\nu_{\text{indir}}}G_s
        &, \text{farside},\\
    \end{cases}
\end{align}
where,
\begin{align}
    \nu_{\text{indir}}
    =\frac{1}{2}\left(\text{sign}[\cos(\theta_{\rm o})]+\text{sign}[\cos(\theta_s)]\right)
\end{align}

An implication of \eqref{eqn:Bardeen_coordinates} is that $\text{sign}(\beta)=\left.\text{sign}(p_\theta)\right\rvert_{r=\infty}$ of a photon received by an asymptotic observer.
We will use this fact to classify trajectories seen by IC observers as either `far side' or `near side' through the label,
\begin{align}
    \text{isfarside}
    =
        (\text{sign}(\beta)>0)\veebar(\cos\theta_{\rm o} < 0)
\end{align}

We can define $\left.\Delta\tau\right\rvert_{\text{OC}}$ in a similar way to $\left.\Delta\tau\right\rvert_{\text{IC}}$.
The labeling of emission as farside vs. nearside is however implicit since $\text{Im}(\gamma_\phi)$ lies to one side of theta $\beta=0$, and thus will depend on $\left.\text{sign}(p_\theta)\right\rvert_{r=r_s}$.
We therefore define $\left.\Delta\right\rvert_{\text{OC}}$ as:
\begin{align}
    \left.\Delta\tau_0\right\rvert_{\text{OC}}=
    \begin{cases}
        G_s+(-1)^{\nu_{\text{indir}}}G_o&, \text{nearside}\\
        \hat G - G_s+(-1)^{\nu_{\text{indir}}}G_o
        &, \text{farside},\\
    \end{cases}
\end{align}
whose form we would like to emphasize is similar to $\left.\Delta\tau_0\right\rvert_{\text{IC}}$, but with $G_s$ and $G_o$ swapped.
Thus we have that,
\begin{align}
    \Delta\tau_0 = 
    \begin{cases}
        \left.\Delta\tau_0\right\rvert_{\text{IC}}&,|\cos\theta_{\rm o}| > |\cos\theta_s|\\
        \left.\Delta\tau_0\right\rvert_{\text{OC}}&,|\cos\theta_{\rm o}| < |\cos\theta_s|
    \end{cases}\label{eqn:minotime_0}
\end{align}

Figure \ref{fig:0-indexing} shows a summary of the classification of $\Delta\tau_0$ trajectories.

\eqref{eqn:minotime_0} can be further simplified with Boolean algebra. 
We define the $\nu_\theta$ at the points of emission as,
\begin{align}
    \nu_\theta
        =\text{sign}(p_\theta(\theta_s))
        =
    \begin{cases}
        (-1)^{n + \text{sign}(\theta_{\rm o}-\theta_s)+1},
        & \text{IC} 
        \\
        (-1)^{\text{sign}\left(\frac{\pi}{2}-\theta_s\right) + 1 },
        &
        \text{near side \& OC}\\
         (-1)^{\text{sign}\left(\theta_s-\frac{\pi}{2}\right) + 1 },
        &
        \text{far side \& OC}
    \end{cases},
\end{align}
which allows the total minotime accrued to be written as,
\begin{align}
    \Delta\tau_n =
    \begin{cases}
        n\hat{G} - \text{sign}(\beta)G_o+\nu_\theta G_s
        &\text{nearside}\\
        (n+1)\hat{G} - \text{sign}(\beta)G_o+\nu_\theta G_s
        &\text{farside}  
    \end{cases}
\end{align}
where $n$ is the image index.
\section{Image Domain}
\label{sec:image-domain-appendix}
\begin{figure*}
    \centering
    \includegraphics[width=\textwidth]{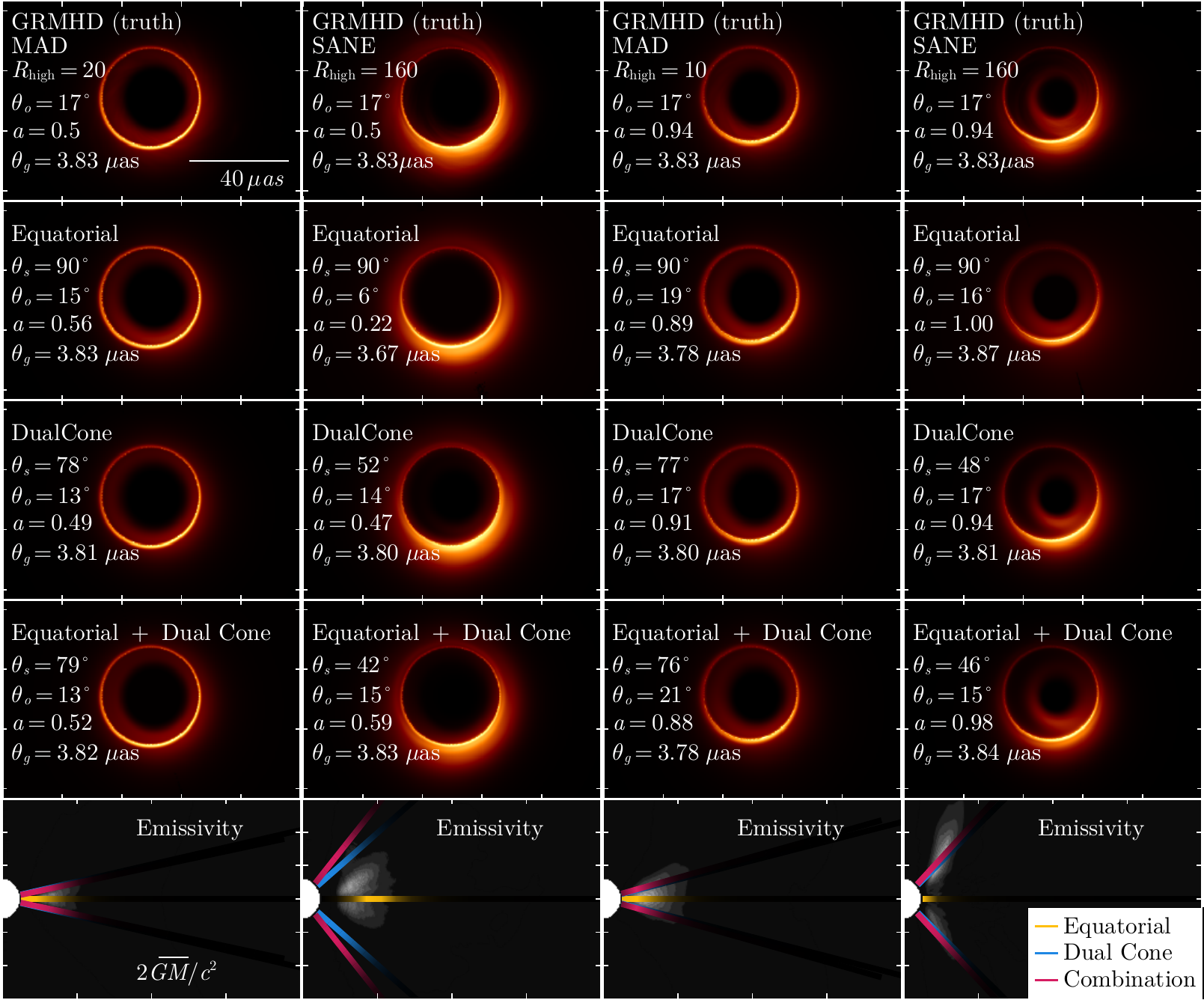}
    \caption{Best-fit (by optimizing the normalized cross correlation; see \autoref{sec:image_domain}) equatorial, dual cone, and combination (equatorial + dual cone) models to four GRMHD simulations of \m87. The four models are (left) $a=0.5$, \rhigh$=20$, MAD, (center-left) $a=0.5$, \rhigh$=160$, SANE, (center-right) $a=0.94$, \rhigh$=10$, MAD, and (right) and $a=0.94$, \rhigh$=160$, SANE. The bottom row compares the model emissivity to the GRMHD emissivity. The equatorial model shows the most significant errors in the best-fit parameters. For the dual cone model, the mass-to-distance ratio of the best-fit model is accurate to within ${\sim}1\%$, the spin is accurate to within 3\%, and the inclination is accurate to within $3^\circ$ for each of the four GRMHD simulations. 
    }
    \label{fig:m87-nxcorr-summary}
\end{figure*}

\autoref{sec:image_domain} discusses the details of our image domain study of our semi-analytic models.
The images analyzed were used to illustrate our models' capabilities for reproducing structures seen in various time-averaged GRMHD simulations, MAD and SANE.
We also study the robustness of our model at producing representative images of GRMHD at various inclinations and at limited resolution. 
We show that the parameters that define the representative images are, in general, consistent with the parameters of the true GRMHD simulations.
\autoref{tab:best_nxcorr} list a summary of the parameters that we used to define our models, along with the parameters that were fitted from GRMHD with our combinationation model in \autoref{fig:nxcorr_summary}.
The GRMHD model used in \autoref{fig:nxcorr_summary} was of a retrograde MAD GRMHD.
We note that the representative models at $30^\circ$ and $50^\circ$ are also described by a retrograde accretion flow (The accretion flow is retrograde if the signs of $\chi$ and $a$ differ in our convention.).
We also show additional images of fits to 0.5 spin MAD and SANE GRMHD simulations in \autoref{fig:m87-nxcorr-summary}, where we find similar results as in \autoref{fig:emissivity}.
All fits shown in \autoref{sec:image_domain} were to high spin images of GRMHD.
\begin{table*}[ht]
\centering
\begin{center}
\setlength\extrarowheight{3pt}
\setlength{\tabcolsep}{3pt}
\begin{tabular}{llcc|cccc} \hline \hline
\textbf{Params.} & \textbf{Description} & \textbf{Units} & \textbf{Range} & \multicolumn{4}{c}{\textbf{Best Fit Params. }} \\ 
\cline{5-8}& & & & $\mathbf{30}^\circ$ & $\mathbf{50}^\circ$ & $\mathbf{70}^\circ$ & $\mathbf{90}^\circ$ \\ 
\hline
$\theta_g$ & Mass-to-distance ratio & $\mu\text{as}$ & $[1.5, 8
]$ & 5.04 & 5.04 & 5.04 & 4.99  \\
$a$ & Black hole dimensionless spin & \dots & $[-1, 0]$  & -0.96 & -0.95 & -0.95 & -0.89 \\
$\theta_{\rm o}$ & Observer inclination& deg & $\left[\theta_t-20, \theta_t+20\right]$ & $27^\circ$ & $45^\circ$ & $55^\circ$ & $70^\circ$\\
$\theta_s$ & Cone opening angle& deg& $\left[20, 90\right]$  & $67^\circ$ & $70^\circ$ & $61^\circ$ & $77^\circ$\\
$p.a.$ & Position angle of projected spin axis& deg& \dots &\dots&\dots&\dots&\dots\\
$R_{\text{cone}}$& Characteristic radius of emission on conical component& $\frac{GM}{c^2}$ & $[1, 10]$  & 3.13 & 4.36 & 5.26 & 4.27\\
$p_1{}_\text{cone}$& Inner exponent of the cone number density function& \dots & $[0.1, 10]$  & 0.79 & 0.27 & 0.15 & 0.43\\
$p_2{}_\text{cone}$& Outer exponent of the cone number density function& \dots & $[1, 10]$  & 3.28 & 3.57 & 3.94 & 5.56\\
$\chi_\text{cone}$& Fluid velocity azimuthal angle in ZAMO frame & deg & $[-180, 180]$  & $180^\circ$ & $180^\circ$ & $100^\circ$ & $105^\circ$\\
$\iota_\text{cone}$& Orthogonal angle of magnetic field conical component & deg & $[0, 90]$  & $64^\circ$ & $90^\circ$ & $65^\circ$  & $88^\circ$\\
$\beta_v{}_\text{cone}$& Fluid speed of conical component in ZAMO frame & $c$ & $[0, 0.9]$ & 0.38 & 0.18 & 0.10 & 0.16 \\
$\sigma_\text{cone}$& Spectral index of conical component & \dots & $[-1, 3]$ & -0.01 & -0.19 & 1.14 & 2.37\\
$\eta_\text{cone}$& Tangential angle of magnetic field conical component & deg & $[-180,180]$  & $162^\circ$ & $10^\circ$ & $175^\circ$ & $-7^\circ$\\
$R_\text{disk}$ & Characteristic radius of emission on equatorial component& $\frac{GM}{c^2}$ & $[1, 10]$ & 5.36 & 5.67 & 4.45 & 1.01\\
$p_1{}_\text{disk}$& Inner exponent of the disk number density function& \dots & $[0.1, 10]$ & 0.10 & 0.1 & 0.10 & 9.43\\
$p_2{}_\text{disk}$& Outer exponent of the disk number density function& \dots & $[1, 10]$  & 4.47 & 7.12 & 5.63 & 1.55\\
$\chi_\text{disk}$& Fluid velocity azimuthal angle in ZAMO frame & deg & $[-180, 180]$  & $84^\circ$ & $79^\circ$ & $129^\circ$ & $148^\circ$\\
$\iota_\text{disk}$& Orthogonal angle of Magnetic field equatorial component & deg & $[0, 90]$  & $79^\circ$ & $90^\circ$ & $80^\circ$ & $82^\circ$\\
$\beta_v{}_\text{disk}$& Fluid speed of equatorial component in ZAMO frame & $c$ & $[0, 0.9]$  & 0.17 & 0.41 & 0.19 & $0.09$\\
$\sigma_\text{disk}$& Spectral index of equatorial component & \dots & $[-1, 3]$  & 1.98 & 3.00 & 1.78 & -1.11\\
$\eta_{disk}$& Tangential angle of magnetic field equatorial component & deg & $[-180,180]$  & $10^\circ$ & $-19^\circ$ & $180^\circ$ & $90^\circ$\\
$r_J$& Relative flux component between cone and disk & \dots& $[0,1]$  & 0.5 & 0.48 & 0.81 & 0.58\\
\hline
\end{tabular}
\caption{ Summary of results from the NxCORR fit study of \autoref{fig:nxcorr_summary}.
The parameters and the search ranges shown include all the parameters that were used in the image domain study of \autoref{sec:image_domain} across all model types. 
$\theta_t$ in the above table refers to the true spin inclination that our model was fitted to.
There are no entries for the $p.a.$ under the `Best Fit Parameters' column since it was fixed to the true value during the fits.
}
\label{tab:best_nxcorr}
\end{center}
\end{table*}

\section{Visibility Domain}
Here we show the full corner plots for the visibility domain fits of \autoref{sec:visibility_domain}.
These include the corner plots for the self fit, the GRMHD fit and the 2017 \m87 data fit.
\begin{figure*}[htbp]
    \centering
    \includegraphics[width=\textwidth]{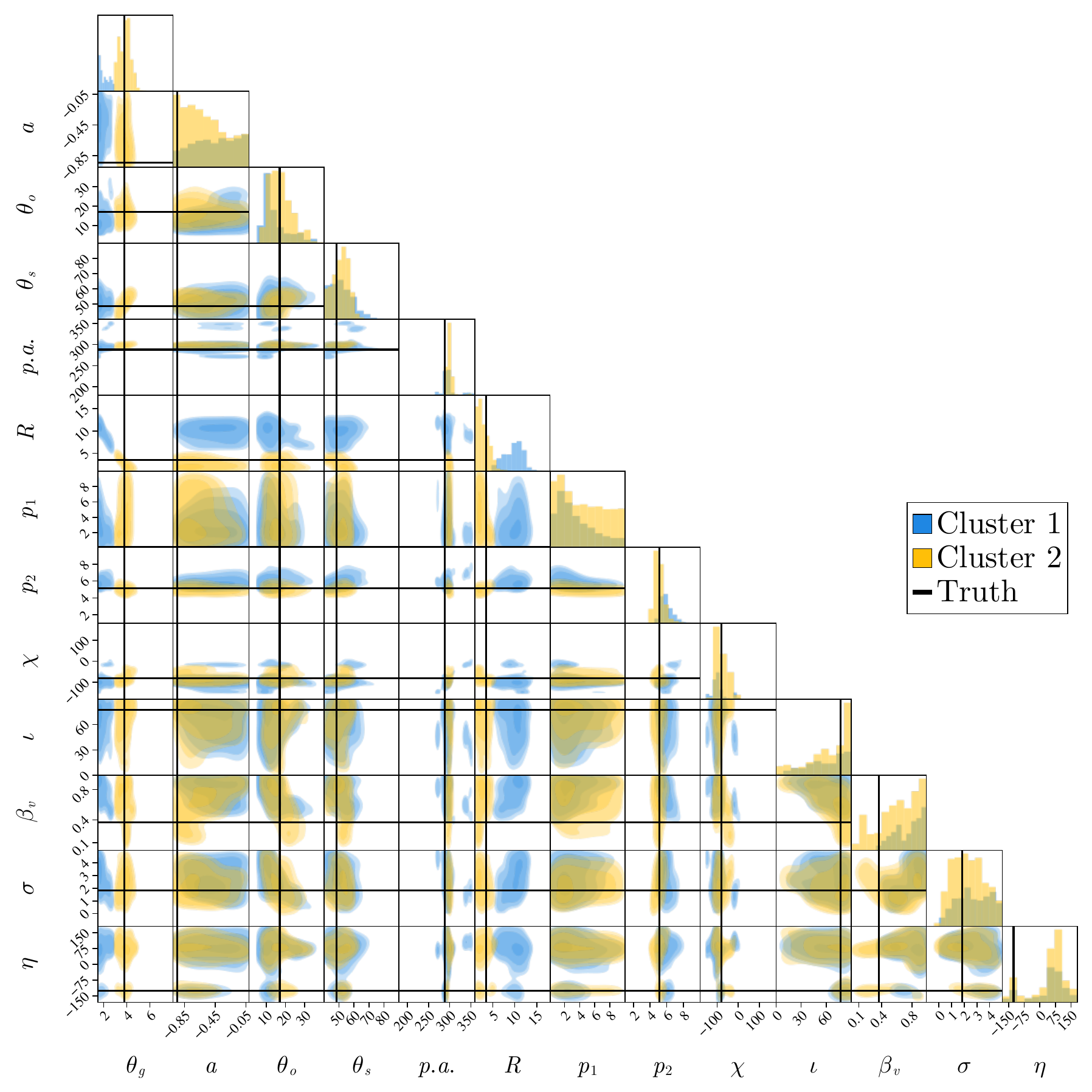}
    \caption{Full triangle plot for dual-cone model fits to synthetic data generated from a dual-cone model (see \autoref{sec:visibility-domain-self-fit} for details).}
    \label{fig:visibility-self-fit-full}
\end{figure*}
\begin{figure*}[htbp]
    \centering
    \includegraphics[width=\textwidth]{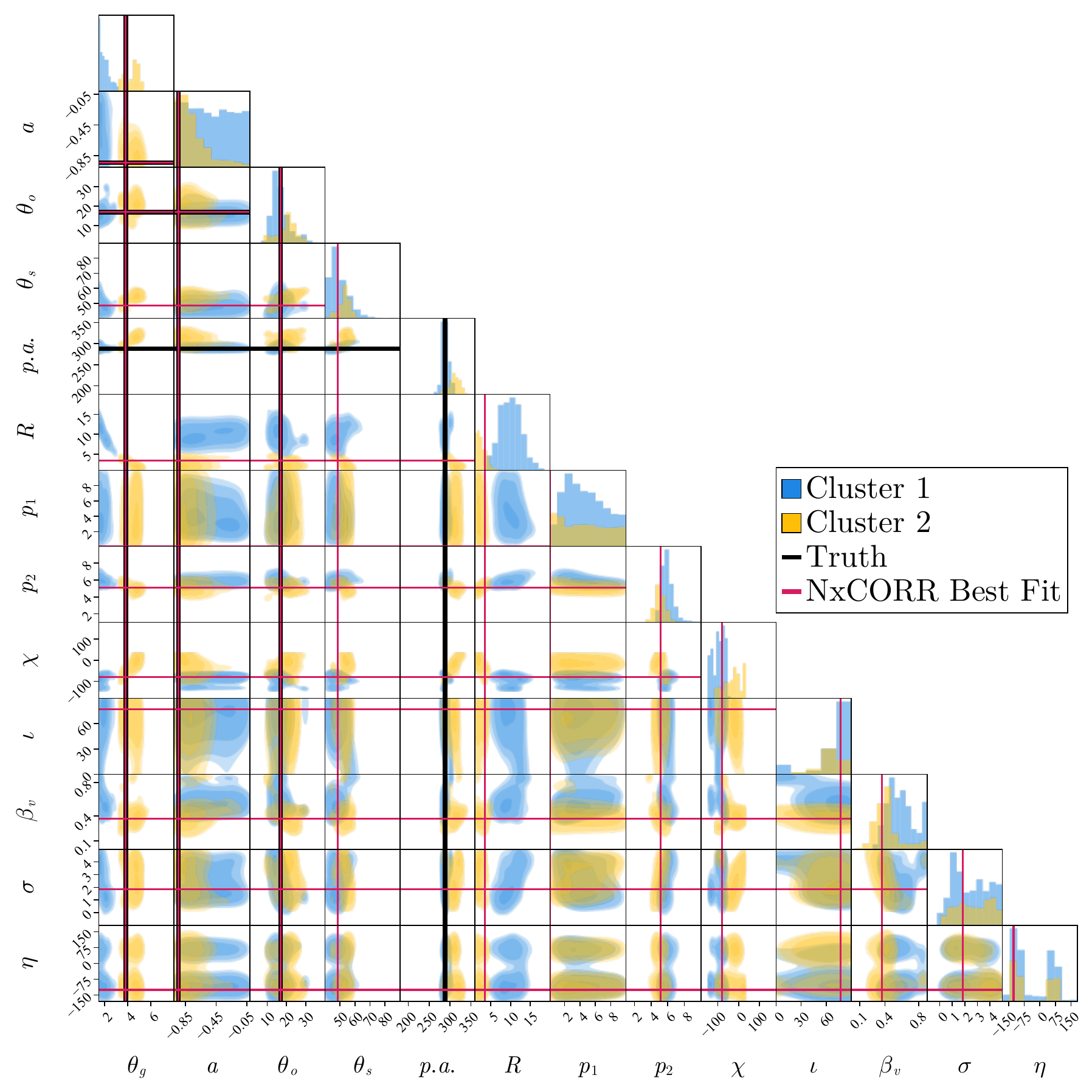}
    \caption{Full triangle plot for dual-cone model fits to synthetic data generated from a prograde \rhigh$=160$ SANE simulation of \m87 (see \autoref{sec:visibility-domain-grmhd-fit} for details).}
    \label{fig:visibility-grhmd-fit-full}
\end{figure*}
\begin{figure*}[htbp]
    \centering
    \includegraphics[width=\textwidth]{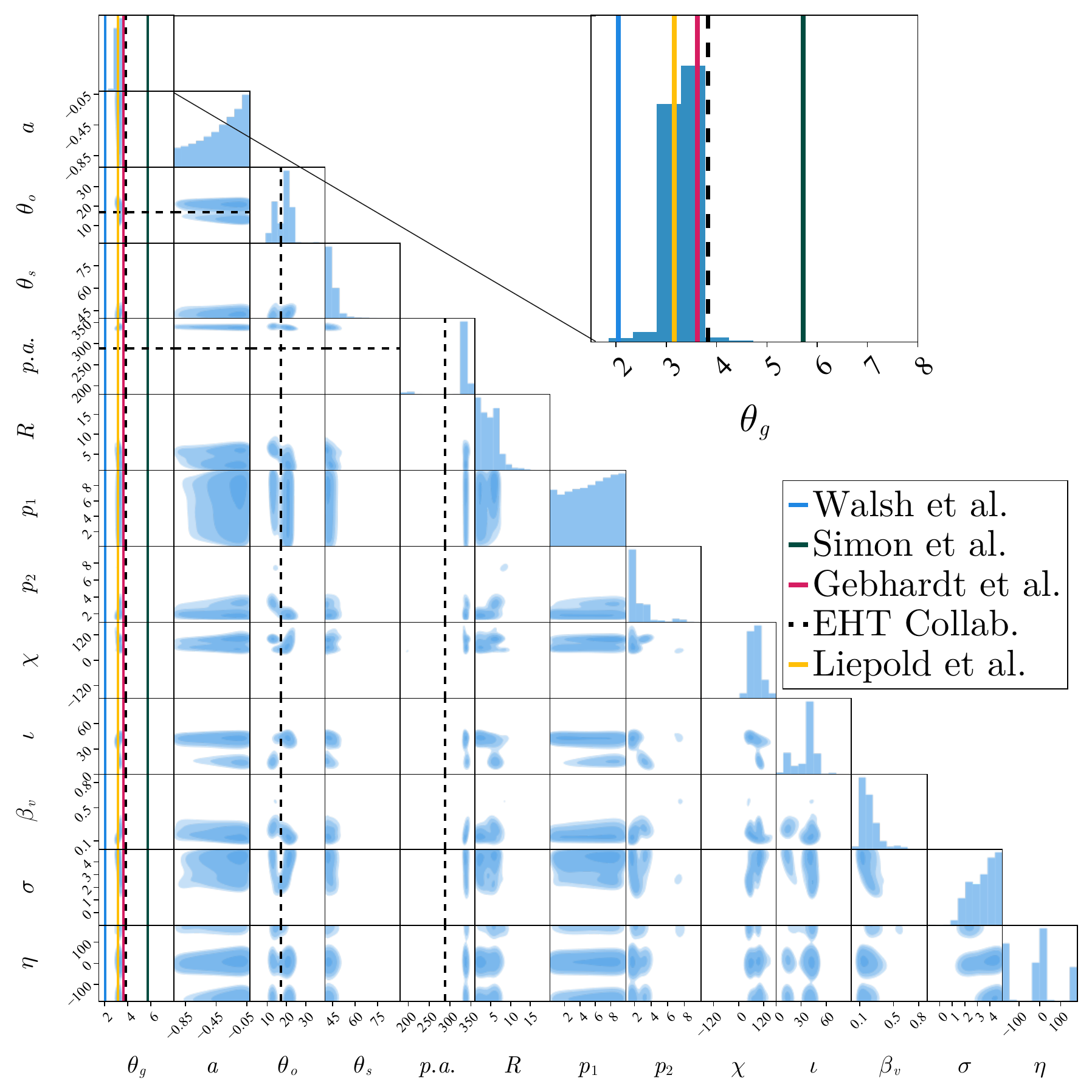}
    \caption{Full triangle plot for dual-cone model fits to EHT measurements of \m87 (see \autoref{sec:visibility-domain-data-fit} for details).}
    \label{fig:visibility-data-fit-full}
\end{figure*}
\bibliography{references}
\end{document}